\documentclass[onecolumn, 11pt, showpacs,preprintnumbers,amsmath,amssymb]{revtex4-1}

\usepackage{graphicx}
\usepackage{dcolumn}
\usepackage{bm}
\usepackage{color,xcolor}

\begin{document}

\title{Evidences for interaction-induced Haldane fractional exclusion statistics in one and higher dimensions}

\author{Xibo Zhang$^{1,2,3}$}
\email{Corresponding author. Email: xibo@pku.edu.cn}
\author{Yang-Yang Chen$^{4,5}$}
\author{Longxiang Liu$^{6,7}$}
\author{Youjin Deng$^{6,7}$}
\author{Xiwen Guan$^{4,8}$}
\affiliation{$^1$International Center for Quantum Materials, School of Physics, Peking University, Beijing 100871, China}
\affiliation{$^2$Collaborative Innovation Center of Quantum Matter, Beijing 100871, China}
\affiliation{$^3$Beijing Academy of Quantum Information Sciences, Beijing 100193, China}

\affiliation{$^4$State Key Laboratory of Magnetic Resonance and Atomic and Molecular Physics, WIPM, Innovation Academy for Precision Measurement Science and Technology, Chinese Academy of Sciences, Wuhan 430071, China}

\affiliation{$^5$Shenzhen Institute for Quantum Science and Engineering, and Department of Physics, Southern University of Science and Technology, Shenzhen 518055, China}
\affiliation{$^6$Shanghai Branch, National Research Center for Physical Sciences at Microscale and Department of Modern Physics, University of Science and Technology of China, Shanghai 201315, China}
\affiliation{$^7$Chinese Academy of Sciences Center for Excellence: Quantum Information and Quantum Physics, University of Science and Technology of China, Hefei Anhui 230326, China}
\affiliation{$^8$Department of Theoretical Physics, Research School of Physics and Engineering, Australian National University, Canberra ACT 0200, Australia}


\begin{abstract}
Haldane fractional exclusion statistics (FES) has a long history of intense studies, but its realization in physical systems is rare. Here we study repulsively interacting Bose gases at and near a quantum critical point, and find evidences that such strongly correlated gases obey simple  non-mutual FES over a wide range of interaction strengths in both one and two dimensions.	Based on exact solutions in one dimension, quantum Monte Carlo simulations and experiments in both dimensions, we show that the thermodynamic properties of these interacting gases, including entropy per particle, density and pressure, are essentially equivalent to those of non-interacting particles with FES. Accordingly, we establish a simple interaction-to-FES mapping that reveals the statistical nature of particle-hole symmetry breaking induced by interaction in such quantum many-body systems. Whereas strongly interacting Bose gases reach full fermionization in one dimension, they exhibit incomplete fermionization in two dimensions. Our results open a route to understanding  correlated interacting  systems via non-interacting particles with FES in arbitrary dimensions.

\end{abstract}


\maketitle

\begin{figure}[t]
	\includegraphics[width = 12cm]{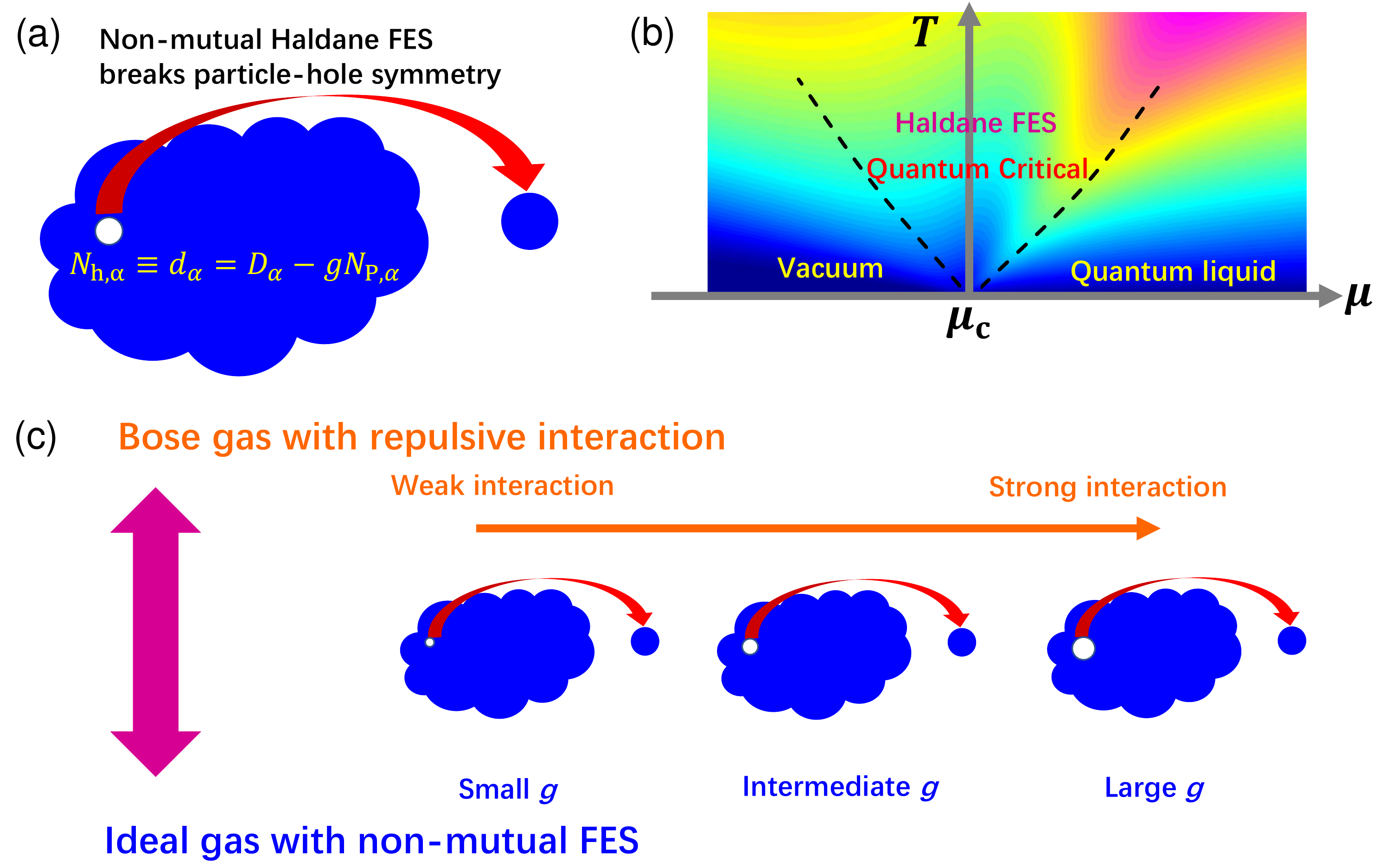}
	\caption{\label{fig1} Schematics of the Haldane FES and its relation to interacting systems. (a) Non-mutual FES. The breaking of particle-hole symmetry is characterized by a single parameter $g$, where the particle number, $N_{\mathrm{P},\alpha}$, in species $\alpha$ only affects the number of holes, $N_{\mathrm{h},\alpha}$, in the same species.  (b) Interacting Bose gases in the quantum critical regime near a vacuum-to-quantum-liquid phase transition. (c) Relation between the interaction strength and $g$: a weaker interaction strength leads to a larger degree of broken particle-hole symmetry.}
\end{figure}

 Bose-Einstein and Fermi-Dirac statistics constitute two cornerstones of quantum statistical mechanics. However, they are not the only possible forms of quantum statistics~\cite{Khare2005FSbook}. In two dimensions (2D), anyonic excitations can carry fractional charges and obey fractional  statistics~\cite{Leinaas1977, Wilczek1982prl48, Wilczek1982prl49, Wilczek84prl, Laughlin88prl, Laughlin88science}. To extend the concept of fractional statistics, Haldane generalized the Pauli exclusion principle and formulated a theory of fractional exclusion statistics (FES) that continuously interpolates between the Bose and Fermi statistics in arbitrary spatial dimensions~\cite{Haldane91PRL}. This theory breaks particle-hole symmetry~\cite{Ha94prl} and defines a FES parameter $g_{\alpha\beta}$ by counting how much the dimensionality $d_{\alpha}$ of the Hilbert space of available single-particle states, namely the ``number of holes'' $N_{\mathrm{h},\alpha}$ of species $\alpha$ decreases as particles of various species $\beta$ are added to a system of fixed size and boundary conditions:
\begin{eqnarray}\label{HaldaneFES}
\Delta N_{\mathrm{h},\alpha} \equiv \Delta d_{\alpha} & = & -\sum_{\beta} g_{\alpha \beta} \Delta N_{\mathrm{P},\beta},
\end{eqnarray}
where $\alpha$ and $\beta$ are ``labels of species'' consisting of a certain set of quantum numbers (such as the quasi-momentum),  $N_{\mathrm{P},\beta}$ is the particle number in species $\beta$, and $g_{\alpha\beta}$ is independent of the particle  numbers. Fig.~\ref{fig1}(a) illustrates a simplified example of non-mutual FES with $g_{\alpha\beta} = g \delta_{\alpha\beta}$ and $D_{\alpha} = \mathrm{max}\{d_{\alpha}\}$, where Bose and Fermi statistics correspond to $g = 0$ and $1$, respectively. The statistical distribution of particles in an ideal gas with FES can be derived via the standard methods in statistical mechanics~\cite{Wu94PRL, Isakov94PRL}.

Haldane's FES approach reveals the statistical nature of a physical system with respect to its energy spectrum rather than the exchange statistics of wave functions. Therefore, it applies to generic quantum matters regardless of whether the constituent particles are interacting or not. As a result, FES provides a powerful approach for studying interacting quantum many-body systems, and has found realizations in a few physical systems. In one dimension (1D), the Calogero-Sutherland model of particles interacting through a  $1/r^2$ potential~\cite{Calogero69JMP2191, Calogero69JMP2197, Sutherland71JMP, Wu94note}, Lieb-Liniger Bose gases~\cite{LiebLiniger63pr, Wu94note}, and anyonic gases with delta-function interaction~\cite{Kundu99prl,Guan06prl} have been exactly mapped onto ideal gases with FES. In three dimensions, FES was assumed to be valid and used to analyze the equation of states of unitary Fermi gases~\cite{Bhaduri07JPB}. On the other hand, it remains challenging to find evidences for FES in generic interacting quantum systems with varied spatial dimensionality and interaction strength.

In this letter, we consider repulsively interacting Bose gases at and near a quantum critical point in 1D and 2D. Under zero temperature, a system undergoes a quantum phase transition from a vacuum to a quantum liquid when the chemical potential $\mu$ exceeds a critical value $\mu_{\mathrm{c}}$ (Fig.~\ref{fig1}(b)). Here ``quantum liquid''  denotes Tomonaga-Luttinger liquid (TLL)~\cite{GuanYuan17PRL} in 1D or superfluid in 2D~\cite{Zhang12Science}. We study non-mutual Haldane FES in such systems and show the relation between the interaction strength and FES parameter $g$ (Fig.~\ref{fig1}(c)).

\begin{figure}[t]
	\includegraphics[width = 12cm]{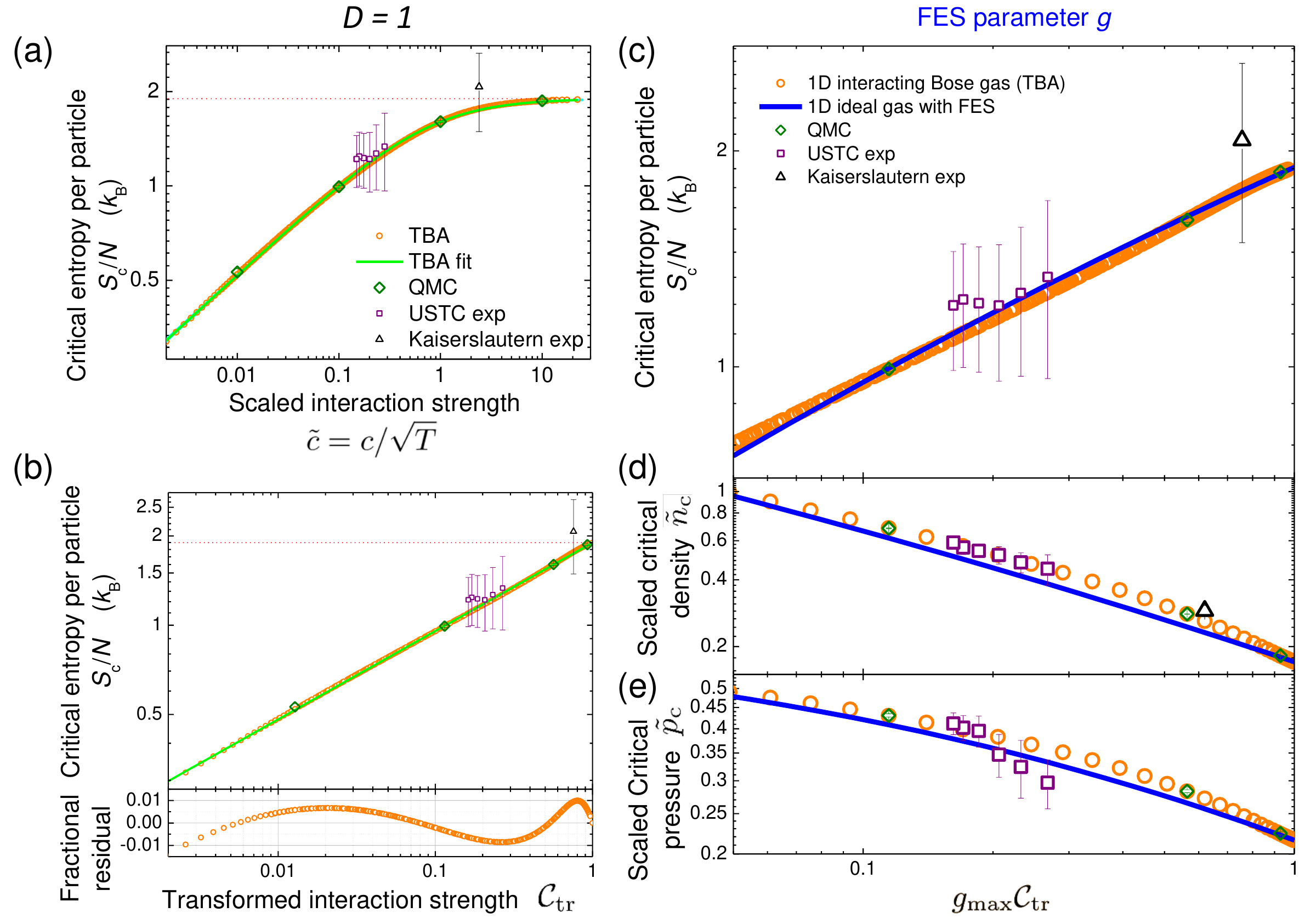}
	\caption{\label{fig2} Evidences for interaction-induced FES in 1D Bose gases at a quantum critical point. (a) Critical entropy per particle $S_{\mathrm{c}}/N$ as a function of the scaled interaction strength $\tilde{c}$. Exact solutions (circles) agree excellently with QMC computations (diamonds) and agree with experiments (squares and triangles, from Refs.~\cite{Ott13PRA,GuanYuan17PRL}). (b) Power-law scaling of $S_{\mathrm{c}}/N$ with respect to $\mathcal{C}_{\mathrm{tr}}$. Dotted line denotes the fermionization limit $A_{\infty, \mathrm{1D}}$.	(c), (d), (e) Evidences for equivalence between a 1D interacting Bose gas and 1D non-interacting particles with non-mutual FES, under the mapping $g = g_{\mathrm{max,1D}}\mathcal{C}_{\mathrm{tr}}$ with $g_{\mathrm{max,1D}} = 1$, regarding three thermodynamic observables: (c) critical entropy per particle $S_{\mathrm{c}}/N$; (d) scaled critical density $\tilde{n}_{\mathrm{c}}$; (e) scaled critical pressure $\tilde{p}_{\mathrm{c}}$.}
\end{figure}

We report evidences for interaction-induced FES in the thermodynamic properties of these Bose gases  based on 
exact solutions in 1D and high-precision quantum Monte Carlo (QMC) simulations  in 2D. Our numerical data are confirmed by existing experiments~\cite{ Ott13PRA, GuanYuan17PRL, Zhang12Science, Dalibard11PRL, Ha13PRL, Hung11nature}. We establish a one-to-one mapping between a transformed interaction parameter  $\mathcal{C}_{\mathrm{tr}}$ and the FES parameter $g$ over a wide range of interaction strengths. Under this mapping, we observe agreements on the entropy per particle, density, and pressure between interacting Bose and non-interacting FES systems at and near the quantum critical point. In 1D, interaction drives the system into a full fermionization with $g_{\mathrm{max,1D}} = 1$, whereas in 2D, $g_{\mathrm{max,2D}} = 0.432(14)$ reveals an incomplete fermionization.

Here we formulate non-interacting particles with 
non-mutual FES parameter $g$. The occupation number $f$ in a state with energy $\epsilon$ is given by~\cite{Wu94PRL} 
\begin{eqnarray}\label{nonmutualFES}
f(\epsilon) & = & \frac{1}{w(\zeta) + g} \nonumber\\
w^g(1+w)^{1-g} & = & \zeta \equiv \exp\left( \frac{\epsilon - \mu}{T} \right) 
\end{eqnarray}
where $T$ is the temperature and $\mu$ the chemical potential. The number density and energy density are given by $n=\int G(\epsilon) f(\epsilon)\mathrm{d}\epsilon$ and $e=\int G(\epsilon) f(\epsilon)\epsilon\mathrm{d}\epsilon$, where the density of states per volume is given by $G(\epsilon) = \frac{1}{2\pi \sqrt{\epsilon}}$ in 1D and $\frac{1}{4\pi}$ in 2D for  non-relativistic particles ($\epsilon = k^2$; $k$ is the momentum). In this work, we set $2 m = k_{\mathrm{B}} = \hbar = 1$, where $m$ is the particle mass, $k_{\mathrm{B}}$ is the Boltzmann constant, and $\hbar$ is the reduced Planck constant. We aim to search for such simple non-mutual FES in strongly correlated matters of ultracold atoms.

The 1D repulsively interacting Bose gases (with no inelastic losses) are described by the Hamiltonian
\begin{eqnarray}\label{Hamiltonian}
\mathcal{H} & = & \sum^N_{i=1}\left(-\nabla^2_i - \mu \right) + c \sum_{i \neq j} \delta(\mathbf{r}_i - \mathbf{r}_j),
\end{eqnarray}
where $c$ is the interaction strength and $N$ is the number of particles. In its dilution limit, the discrete 1D Bose-Hubbard model used in QMC simulations relates to Eq.~\ref{Hamiltonian} via  $c = \frac{U}{2t^{1/2}}$~\cite{SM19Zhang}, where $U$ and $t $ are the onsite interaction and tunneling parameters, respectively.   

We study 1D Bose gases with repulsive delta-function interaction (Eq.~\ref{Hamiltonian})~\cite{LiebLiniger63pr,YangYang69JMP}   at the vacuum-to-TLL transition ($\mu_{\mathrm{c}} =0$)~\cite{GuanYuan17PRL}. This system is exactly solvable via thermodynamic Bethe Ansatz (TBA) equation~\cite{YangYang69JMP}:
\begin{eqnarray}\label{TBA}
\epsilon(k) & = & k^2 - \mu - \frac{T}{2\pi}\int \mathrm{d}p a(k-p)\ln\left(1+e^{-\frac{\epsilon(p)}{T}} \right),
\end{eqnarray}
where $a(q) = 2c / (c^2 + q^2)$, and the pressure is given by $p(\mu,T) = \frac{T}{2\pi}\int\ln\left(1 + e^{-\epsilon(k)/T}\right)\mathrm{d}k$. For convenience in analysis, we present   thermodynamic observables and parameters in scaled dimensionless  forms~\cite{SM19Zhang}. We compute thermodynamic observables such as the critical entropy per particle $S_{\mathrm{c}}/N \equiv \frac{S}{N}\left(\mu = \mu_{\mathrm{c}}\right)$,  scaled critical density $\tilde{n}_{\mathrm{c,1D}}  = n_{\mathrm{c}}/T^{1/2}$, and scaled critical pressure $\tilde{p}_{\mathrm{c,1D}} = p_{\mathrm{c}}/T^{3/2}$ by numerically solving Eq.~\ref{TBA}.

We show  $S_{\mathrm{c}}/N$ increases with the growth of a scaled interaction strength $\tilde{c} = c/\sqrt{T}$ (Fig.~\ref{fig2}(a)).  At $\tilde{c} \rightarrow \infty$, it reaches  $A_{\infty,\mathrm{1D}}  \approx 1.89738$ (dotted line)~\cite{SM19Zhang} that exactly matches the $S_{\mathrm{c}}/N$ of non-interacting  fermions~\cite{YangYang69JMP} (corresponding to $g = 1$).  Our TBA solutions agree with data extracted from experiments performed by the Kaiserslautern group~\cite{Ott13PRA} and the USTC group~\cite{GuanYuan17PRL}, and agree with our 1D QMC simulations~\cite{SM19Zhang}.

We further observe that $S_{\mathrm{c}}/N$ obeys an empirical power-law scaling (Fig.~\ref{fig2}(b))
\begin{eqnarray}\label{powerlaw1D}
\frac{S_{\mathrm{c}}}{N} & = & A_{\infty,\mathrm{1D}} \mathcal{C}_{\mathrm{tr}}^{\beta_{\mathrm{1D}}}
\end{eqnarray}
with respect to a transformed interaction parameter $\mathcal{C}_{\mathrm{tr}}$:
\begin{eqnarray}\label{Ctr}
\mathcal{C}_{\mathrm{tr}} & \equiv & \frac{\tilde{c}/\tilde{c}_1}{\tilde{c}/\tilde{c}_1+1},
\end{eqnarray}
where $\beta_{\mathrm{1D}} = 0.298(1)$ and   $\tilde{c}_{1} = 0.772(5)$ in 1D are determined by a two-parameter fit. The fit agrees with the TBA data within $1\%$ over a large range of interaction strengths ($0.002 < \tilde{c} < \infty$). Equation~\ref{Ctr} is inspired by a Ginzburg-Landau theory~\cite{Sachdev04prb} for 2D superfluid~\cite{Ha13PRL}. The dependence of $S_{\mathrm{c}}/N$ on interaction, observed in a previous experiment~\cite{Zhang12Science}, is accurately described here as a power-law scaling (Eq.~\ref{powerlaw1D}) that accordingly signals interaction-induced FES, as explained below.

We find a simple and explicit interaction-to-FES mapping by comparing Eq.~\ref{powerlaw1D} with the behavior of  non-interacting particles with non-mutual FES. The TBA equation is a consequence of breaking  particle-hole symmetry in excitations determined by the Bethe Ansatz equations~\cite{Guan07laserPL}. Such particle-hole symmetry breaking can in general be quantified by momentum-dependent mutual FES~\cite{Wu94note}, and can be described by non-mutual FES in strongly interacting systems~\cite{ Guan07laserPL}. At and near a quantum critical point where the correlation length is large, the underlying FES physics can be greatly simplified into non-mutual FES. To demonstrate this point, we compute the ``critical'' entropy per particle (at $ \mu_{\mathrm{c}} = 0$), $ S_{\mathrm{c,FES}}/N $, of a non-interacting gas with a non-mutual FES parameter $g$ (Fig.~\ref{fig2}(c), blue curve), and find that $ S_{\mathrm{c,FES}}/N $  exhibits an approximate power-law scaling with respect to $g$: $S_{\mathrm{c,FES}}/N = A_{\infty,\mathrm{1D}} g^{\beta_{\mathrm{FES,1D}}}$, with $\beta_{\mathrm{FES,1D}} = 0.298(2)$ fitted for $0.05 < g \le 1$. This second power-law scaling and Eq.~\ref{powerlaw1D} agree very well, and the corresponding numerical data agree within 4\% (Fig.~\ref{fig2}(c)), which strongly suggests a one-to-one mapping between an interacting Bose gas with  $\mathcal{C}_{\mathrm{tr}}$ and a non-interacting FES gas with $g$:
\begin{eqnarray} \label{mapping}
g & = & g_{\mathrm{max}} \mathcal{C}_{\mathrm{tr}},
\end{eqnarray}
with $g_{\mathrm{max}} = 1$ in 1D.

We further support Eq.~\ref{mapping} by showing similar agreements for  $\tilde{n}_{\mathrm{c}}$ and  $\tilde{p}_{\mathrm{c}}$ (Figs.~\ref{fig2}(d) and \ref{fig2}(e)). The agreements are within $15\%$ for $\tilde{n}_{\mathrm{c}}$ and  $8\%$ for $\tilde{p}_{\mathrm{c}}$. The overall good agreements on  $S_{\mathrm{c}}/N$, $\tilde{n}_{\mathrm{c}}$, and $\tilde{p}_{\mathrm{c}}$ provide a comprehensive set of evidences for interaction-induced FES in a  1D Bose gas at a quantum critical point. Here, $g_{\mathrm{max}} = 1$ corresponds to a full fermionization of a 1D  Bose gas at $\tilde{c} = \infty$, which was predicted and observed for quantum gases in the Tonks-Girardeau regime~\cite{Tonks36pr, Girardeau60jmp, LiebLiniger63pr, Bloch04nature, Weiss04science}.

\begin{figure}[t]
	\includegraphics[width=12cm]{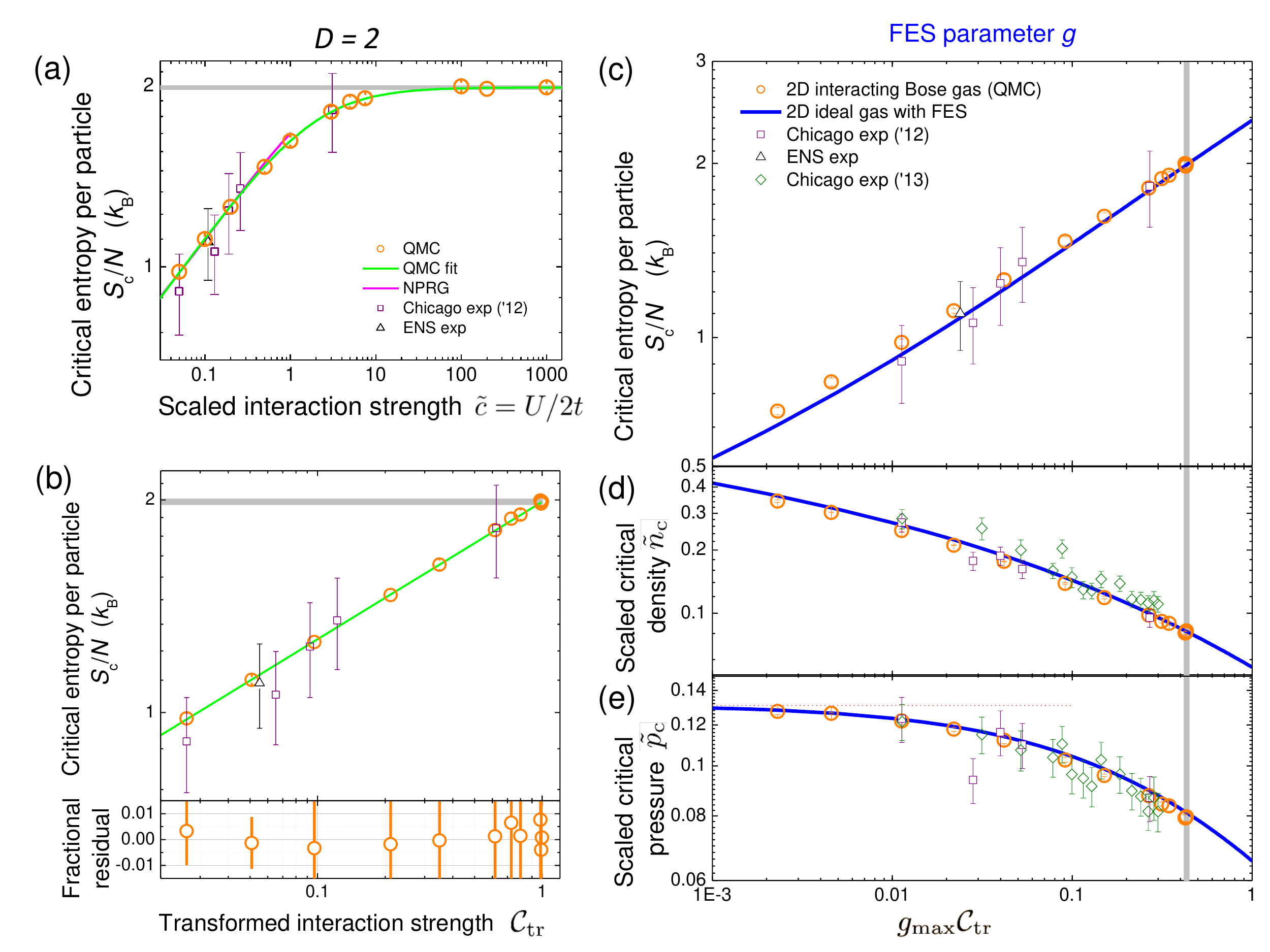}
	\caption{\label{fig3} Evidences for interaction-induced FES in 2D  Bose gases at a quantum critical point. (a) Critical entropy per particle $S_{\mathrm{c}}/N$ as a function of  $\tilde{c}_{\mathrm{2D}}$. QMC results (circles) agree with NPRG  computations~\cite{Rancon12PRA} and  experiments~\cite{Zhang12Science, Dalibard11PRL}.  (b) Power-law scaling of $S_{\mathrm{c}}/N$ with respect to  $\mathcal{C}_{\mathrm{tr}}$. (c), (d), (e) Evidences for equivalence between a 2D interacting Bose gas and 2D non-interacting particles with non-mutual FES,  under the mapping $g = g_{\mathrm{max, 2D}}\mathcal{C}_{\mathrm{tr}}$ with $g_{\mathrm{max,2D}} = 0.432(14)$, regarding three thermodynamic observables: (c) critical entropy per particle $S_{\mathrm{c}}/N$; (d) scaled critical density $\tilde{n}_{\mathrm{c}}$; (e) scaled critical pressure $\tilde{p}_{\mathrm{c}}$, with the dotted line denoting the non-interacting boson limit $\tilde{p}_{\mathrm{c}0} = \frac{\pi}{24}$~\cite{Rancon12PRA}. Our results agree with existing experiments~\cite{Zhang12Science, Dalibard11PRL, Ha13PRL}. Horizontal and vertical gray bands mark $A_{\infty,\mathrm{2D}}$ and $g_{\mathrm{max,2D}}$, respectively.}
\end{figure}

Having established Eq.~\ref{mapping} in 1D, we now investigate whether interaction-induced FES~\cite{Bhaduri96prl, Bhaduri00JPB, Hansson01prl} exists in 2D Bose gases at and near a quantum critical point. Based on QMC simulations~\cite{qmc98pla, qmc98jetp}, we study a 2D Bose-Hubbard lattice gas that has a vacuum-to-superfluid quantum phase transition at $\mu_{\mathrm{c}} = -4t$~\cite{Zhang12Science}. The Bose-Hubbard model with no loss is described by a Hamiltonian:
\begin{small}
\begin{eqnarray}
\hat{H}  =   -t \!\sum_{<i,j>}\! \!\left(\hat{b}_{i}^\dagger\hat{b}_{j}^{}\! +\hat{b}_{j}^\dagger\hat{b}_{i}^{}\right) \!+\! \!\sum_{i}\!\left[\frac{U}{2}\hat{n}_i(\hat{n}_{i}-1)-\mu\hat{n}_i\right], 
\end{eqnarray}
\end{small}
where  $\hat{b}_{i}^\dagger$ and $\hat{b}_{i}$ are the creation and annihilation operators at site $i$, $\hat{n}_{i} = \hat{b}_{i}^\dagger \hat{b}_{i}$, and $<i,j>$ runs over all nearest neighboring sites.  For this 2D lattice gas, we identify a scaled interaction strength $\tilde{c}_{\mathrm{2D}} = U/(2t)$~\cite{SM19Zhang} that is the lattice-gas equivalence~\cite{ Zhang12Science, Ha13PRL} of the interaction parameter $\sqrt{8\pi}a/l_z$ for a weakly interacting 2D Bose gas without lattices, where $a$ is the scattering length  and $l_z$ is an oscillator length~\cite{Hung11nature}.

To obtain physical properties that are insensitive to the lattice structure, we perform QMC simulations for each $\tilde{c}_{\mathrm{2D}}$ at a series of temperatures down to $T = 0.1 t$. We extract scaled thermodynamic quantities $S_{\mathrm{c}}/N$, $\tilde{n}_{\mathrm{c,2D}} = n_{\mathrm{c}}/T$, and $\tilde{p}_{\mathrm{c,2D}} = p_{\mathrm{c}}/T^2$ at each $T$, and then perform extrapolation towards $T=0$ for each  quantity~\cite{SM19Zhang}. We test our extrapolation protocol by studying a 1D Bose-Hubbard system~\cite{SM19Zhang} and find excellent agreements with solutions to the TBA equation  (Fig.~\ref{fig2}).

\begin{figure}[t]
	\includegraphics[width=12cm]{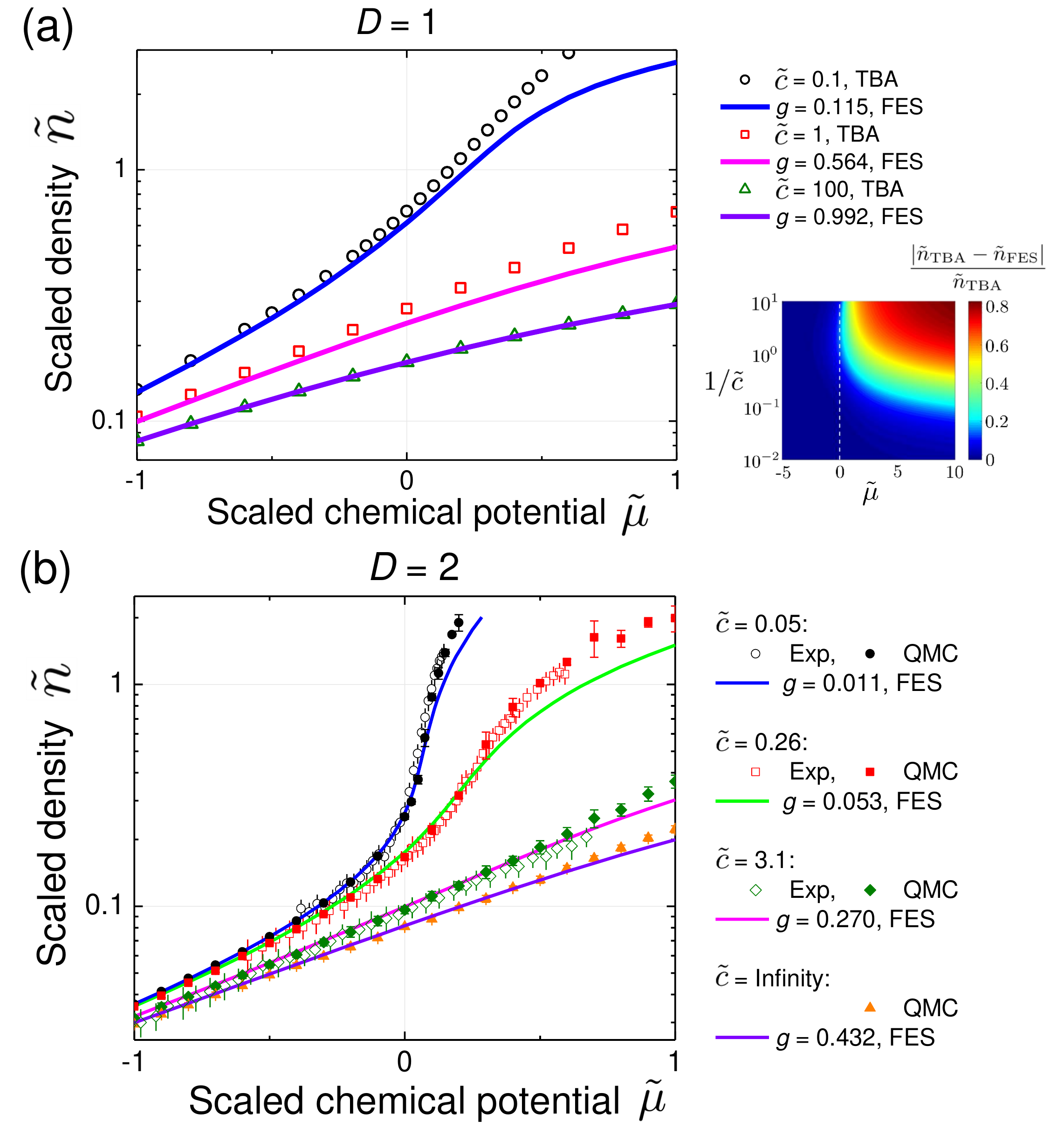}
	\caption{\label{fig4}
		Scope of application of non-mutual FES for describing interacting Bose gases near a quantum critical point. (a) Scaled density $\tilde{n}$ as a function of  $\tilde{\mu} = \frac{\mu - \mu_{\mathrm{c}}}{T}$: 1D interacting Bose gases (open symbols) compared to 1D non-interacting particles with FES (lines). The colored map further illustrates the scope of application of non-mutual FES over various  $\tilde{\mu}$ and $1/\tilde{c}$ values. (b) $\tilde{n}(\tilde{\mu})$ for 2D interacting Bose gases: experimental measurements from Refs.~\cite{Hung11nature, Zhang12Science}(open symbols) and QMC simulations (solid symbols), compared to $\tilde{n}(\tilde{\mu})$ for 2D non-interacting particles with FES (lines). In both (a) and (b), the mappings of $\tilde{c}$ to $g$ are independent of $\tilde{\mu}$ and identical to those used at $\tilde{\mu} = 0$. }
\end{figure}

Based on analyses similar to our 1D studies, we find evidences for interaction-induced FES in 2D Bose gases at the quantum critical point. The $S_{\mathrm{c}}/N$ increases with the growth of $\tilde{c}_{\mathrm{2D}}$ and reaches $A_{\infty,\mathrm{2D}} = 1.988(14)$ at $\tilde{c}_{\mathrm{2D}} = \infty$ (Fig.~\ref{fig3}(a)). Our QMC data for $S_{\mathrm{c}}/N$ agree well with a non-perturbative renormalization group (NPRG) computation (reliable for $\tilde{c}_{\mathrm{2D}} < 1$)~\cite{Rancon12PRA}, and agree with experiments by the Chicago~\cite{Zhang12Science} and ENS~\cite{Dalibard11PRL} groups. Based on Eq.~\ref{Ctr} and $\tilde{c}_{1,\mathrm{2D}} = 1.9(3)$, $S_{\mathrm{c}}/N$ shows an excellent power-law scaling with respect to  $\mathcal{C}_{\mathrm{tr}} = \frac{\tilde{c}_{\mathrm{2D}}/\tilde{c}_{1,\mathrm{2D}}}{\tilde{c}_{\mathrm{2D}}/\tilde{c}_{1,\mathrm{2D}} + 1}$ (Fig.~\ref{fig3}(b)):  $S_{\mathrm{c}}/N = A_{\infty,\mathrm{2D}} \mathcal{C}_{\mathrm{tr}}^{\beta_{\mathrm{2D}}}$, with $A_{\infty,\mathrm{2D}}$, $\tilde{c}_{1,\mathrm{2D}}$, and  $\beta_{\mathrm{2D}} = 0.20(1)$ fitted for $ \tilde{c}_{\mathrm{2D}} \ge 0.05$ ($0.026 \le \mathcal{C}_{\mathrm{tr}} \le 1$). We observe a second power-law scaling, $S_{\mathrm{c,FES}}/N = A_{\mathrm{FES, 2D}} g^{\beta_{\mathrm{FES,2D}}}$, with $A_{\mathrm{FES, 2D}} \equiv \frac{S_{\mathrm{c,FES}}}{N}(g=1) \approx 2.373$. The exponent $\beta_{\mathrm{FES, 2D}} = 0.2122(1)$ is fitted for $0.02 \le g \le 1$ and agrees well with $\beta_{\mathrm{2D}}$, whereas $A_{\infty,\mathrm{2D}}$ is substantially smaller than $A_{\mathrm{FES, 2D}}$. Hence by choosing $g_{\mathrm{max, 2D}} = 0.432(14)$ based on $A_{\infty, \mathrm{2D}}$, we observe that these two power-law scaling functions for $S_{\mathrm{c}}/N$ versus $g_{\mathrm{max, 2D}} \mathcal{C}_{\mathrm{tr}}$ and for $S_{\mathrm{c,FES}}/N$ versus $g$ agree well within $5\%$ (Fig.~\ref{fig3}(c)). Accordingly, $\tilde{n}_{\mathrm{c,2D}}$ and $\tilde{p}_{\mathrm{c,2D}}$ show agreements within $5\%$ and $3\%$, respectively (Figs.~\ref{fig3}(d) and \ref{fig3}(e)). Our numerical results agree with existing experimental measurements~\cite{Zhang12Science, Dalibard11PRL, Ha13PRL},  provide evidences for interaction-induced FES in 2D, and again support a simple interaction-to-FES mapping (Eq.~\ref{mapping}) with a less-than-unity $g_{\mathrm{max, 2D}} = 0.432(14)$. Here, $g_{\mathrm{max, 2D}}$ demonstrates incomplete fermionization of 2D Bose gases at the critical point. We remark that our QMC data  are obtained by spending about $2\times10^5$ CPU hours. It is surprising how well these data are described by the non-mutual FES model whose computation costs only tens of seconds.

To provide a statistical interpretation for Eqs.~\ref{Ctr} and \ref{mapping} based on particle-hole symmetry breaking analysis~\cite{Ha94prl}, we rewrite these two equations as follows:
\begin{eqnarray}\label{statisticalnature}
\frac{g}{ g_{\mathrm{max}}- g} = \frac{\tilde{c}}{\tilde{c}_1}.
\end{eqnarray} 
On the left hand side, the numerator equals the dimensionality of Hilbert space occupied by one single particle; the denominator equals the dimensionality of Hilbert space that is ``unoccupied under  interaction strength $\tilde{c}$'' but still ``occupiable under infinite interaction'' by one single particle. Our work  empirically validates Eqs.~\ref{Ctr} and \ref{mapping} for both 1D and 2D systems at and near a quantum critical point, and hence reveals a phenomenological relation that the dimensionality ratio of the above occupied / unoccupied-but-occupiable Hilbert space is proportional to the scaled interaction strength $\tilde{c}$.

Finally, we further explore the scope of application of our non-mutual FES mapping formula, Eq.~\ref{mapping}, by comparing the scaled equations of state $\tilde{n}(\tilde{\mu}) =\tilde{n}\left(\frac{\mu-\mu_{\mathrm{c}}}{T}\right)$ of interacting Bose gases with those of non-interacting particles with FES in a finite range of $\tilde{\mu}$ besides the quantum critical point $\tilde{\mu}_{\mathrm{c}}$. With no additional adjustable parameters, a strongly interacting 1D Bose gas with $\tilde{c}_{\mathrm{1D}} = 100$ shows excellent equivalence to 1D non-interacting particles with $g = 0.992$ (Fig.~\ref{fig4}(a)). As interaction weakens, the equivalence at $\mu \leq \mu_{\mathrm{c}}$ is still good, whereas deviations become more significant as $\tilde{\mu}$ exceeds $0$. Here we present $\tilde{n}$ because under the same $\tilde{c}$ and $\tilde{\mu}$, the relative deviations for $\tilde{p}$ and $S/N$ are smaller than that for $\tilde{n}$~\cite{SM19Zhang}. Fig.~\ref{fig4}(b) shows similar condition of equivalence between 2D interacting Bose gases and 2D ideal gases with FES. We attribute the deviations at positive finite $\tilde{\mu}$ (Fig.~\ref{fig4}), as well as the residual small deviations for $\tilde{\mu} \le 0$ (Figs.~\ref{fig2}~$\sim$~\ref{fig4}), primarily to the need of including more complex mutual FES effects~\cite{Wu94note, SM19Zhang}, which is subject to future research.

To conclude, we find strong evidences for interaction-induced non-mutual Haldane fractional exclusion statistics in both 1D and 2D Bose gases at and near the vacuum-to-quantum-liquid transition.  Our unified, non-perturbative mapping approach can be generalized by including mutual FES effects and holds promise for providing new insights into interacting fermions for which QMC simulations are in general challenging, and into strongly interacting quantum materials in higher dimensions where experiments can be both enriched and complicated by inelastic collisional losses, three-body effects, finite temperature effects, and the possible break-down of scale invariance~\cite{Ha13PRL, FeiZhou2013prl,Zoran2013prl, CornellJin14NatPhys, Chin16prx, Jochim18prl}.

\begin{acknowledgments}
We are grateful to Cheng Chin for insightful discussions. We acknowledge Chun-Jiong Huang,  Zhen-Sheng Yuan,   Chen-Lung Hung, Li-Chung Ha, Biao Wu, Hui Zhai for discussions and technical support. This work is supported by the National Key Research and Development Program of China under Grant Nos 
2018YFA0305601, 2016YFA0300901, 2017YFA0304500, and 2016YFA0301604, the National Natural Science Foundation of China under Grant Nos 11874073, 11534014, 11874393, 11625522. X.Z., Y.D.  and X.G. conceived the project and wrote the paper.  Y.-Y.C. and L.L. performed the numerical and analytical studies on the 1D and 2D models, respectively. X.Z. also performed analytical and numerical computations for this paper. 
\end{acknowledgments}

\section*{Supplemental material}
\appendix

\section{Yang-Yang thermodynamic Bethe ansatz equation}
The thermodynamic properties of $\delta$-function  interacting Bose gases in one dimension (1D)  can be obtained by solving Yang-Yang thermodynamic Bethe ansatz equation (TBAE)\cite{YangYang69JMP}
\begin{equation}\label{eq:Yang-Yang}
\varepsilon(k)=\frac{\hbar^2 k^2}{2m}-\mu-\frac{k_B T}{2\pi}\int_{-\infty}^{\infty}\frac{2c}{c^2+(k-q)^2}\ln\left(1+e^{-\frac{\varepsilon(q)}{k_B T}}\right) d q,
\end{equation}
where $\varepsilon(k)$ is called ``dressed energy'', $k$ is quasi-momentum, $\mu$ is chemical potential, $T$ is temperature and $c=-2/a_{\rm 1D}$ with $a_{\rm 1D}$ being 1D scattering length. Thus the grand thermodynamic potential of unit length, namely, pressure $p$ can be obtained by
\begin{equation}
p=\frac{k_B T}{2\pi}\int_{-\infty}^{\infty}\ln\left(1+e^{-\frac{\varepsilon(k)}{k_B T}} \right)d k.
\end{equation} 
Other thermodynamic properties, such as particle density $n$, entropy density $s$ are given by the derivatives of the pressure, namely,
\begin{equation}
n=\frac{\partial p}{\partial \mu}\vert_{c,T},\quad s=\frac{\partial p}{\partial T}\vert_{c,\mu}
\end{equation}

\begin{figure}
	\centering
	\includegraphics[scale=0.8]{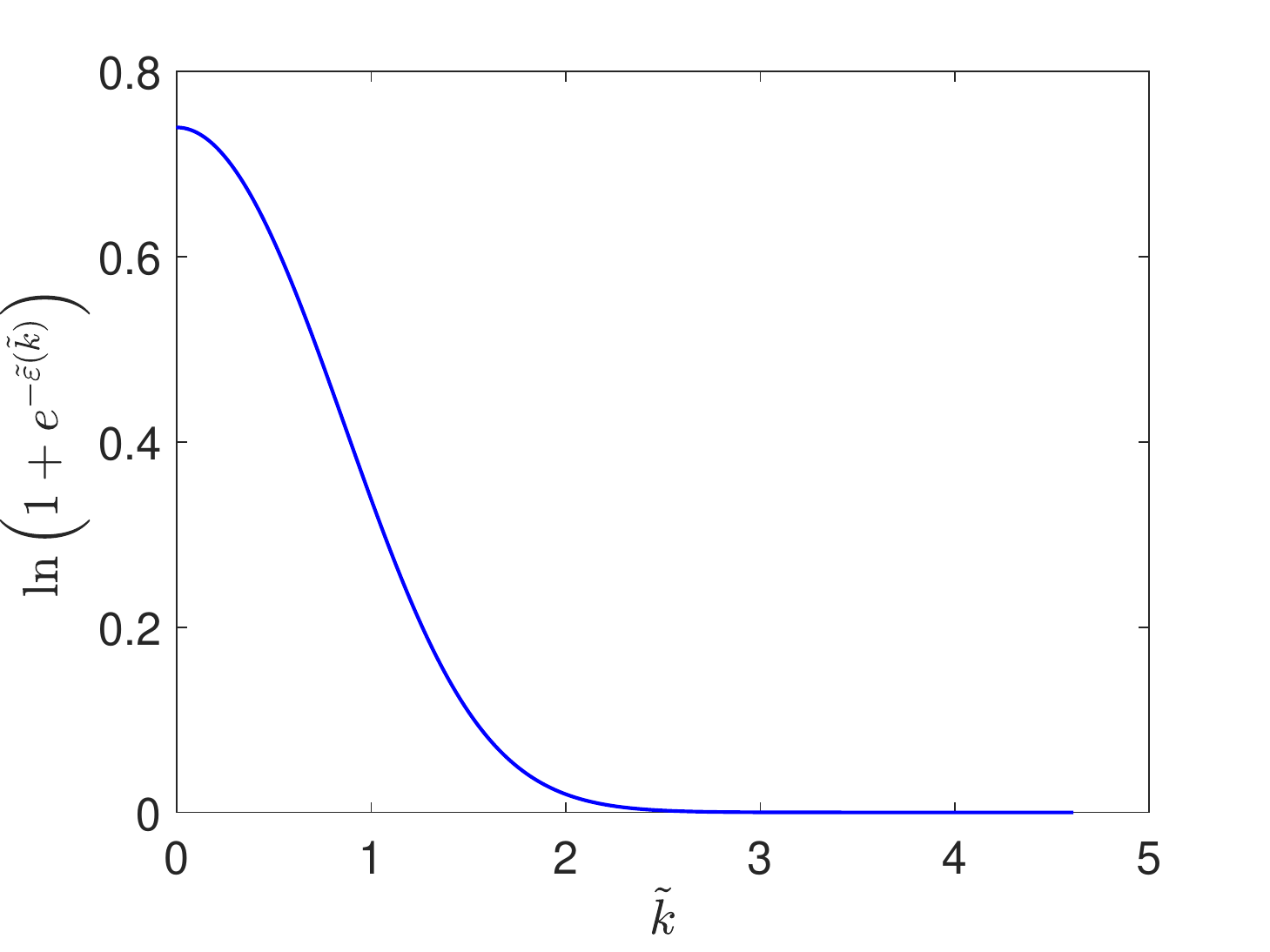}
	\caption{$\ln\left(1+e^{-\tilde{\varepsilon}(\tilde{k})}\right)$ vs $\tilde{k}$. Here we take $\tilde{\mu}=0,\tilde{c}=5$ and the cutoff $\tilde{k}_c\approx 4.6$.}
	\label{fig:S1}
\end{figure}

\begin{figure}
	\centering
	\includegraphics[scale=0.8]{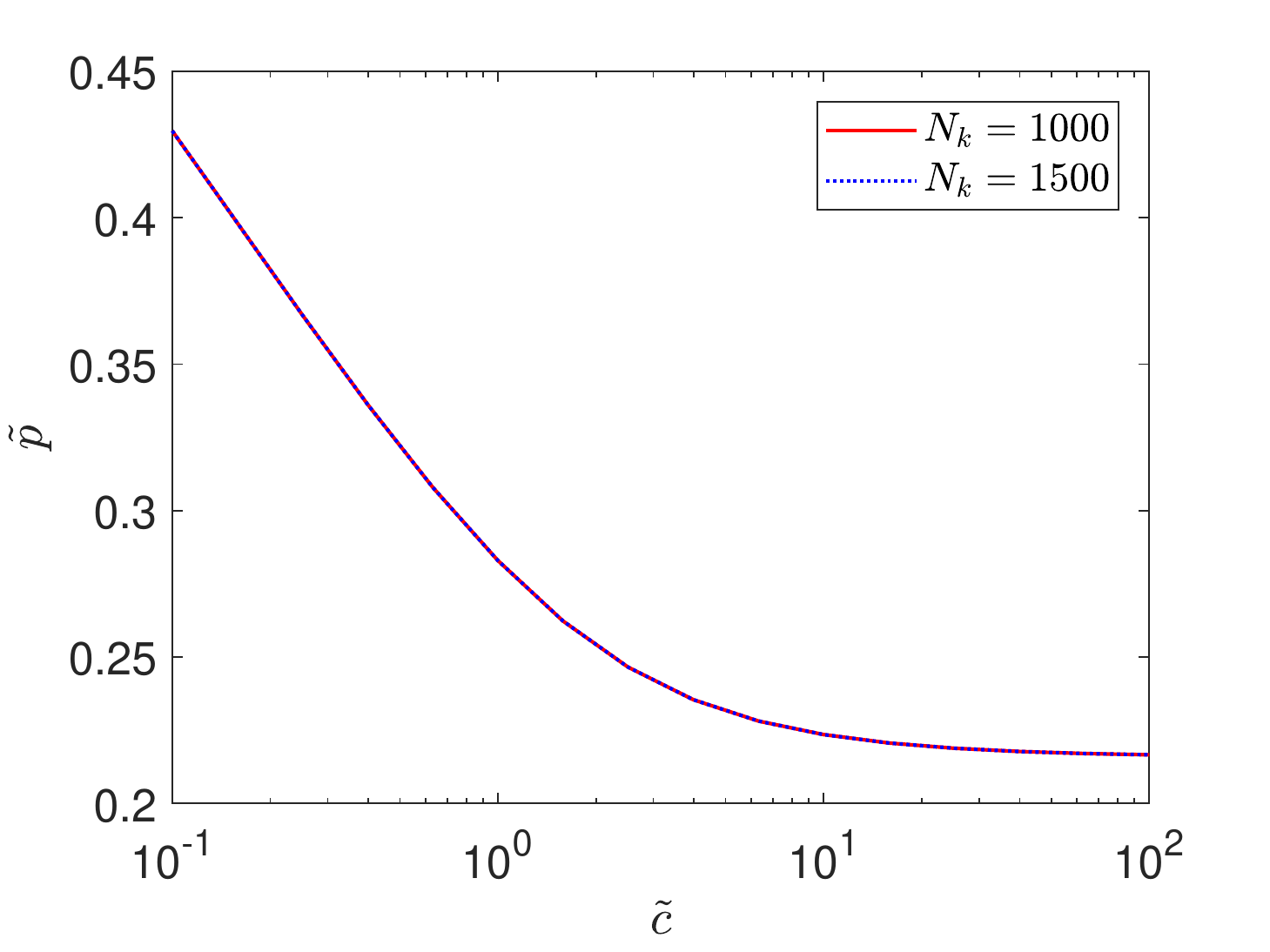}
	\caption{The dimensionless pressure $\tilde{p}$ vs $\tilde{c}$. The red solid line corresponds to the number of discretization $N_k=1000$ and the blue dot line corresponds to $N_k=1500$.}
	\label{fig:S2}
\end{figure}

For our convenience in analysis of critical phenomenon, we define the following dimensionless parameters
\begin{equation}
\tilde{k}\equiv \frac{\hbar k}{\sqrt{2m k_B T}},\quad \tilde{c}\equiv \frac{\hbar c}{\sqrt{2m k_B T}},\quad \tilde{\mu}=\frac{\mu}{k_B T},\quad \tilde{\varepsilon}\equiv \frac{\varepsilon}{k_B T}
\end{equation} and the dimensionless thermodynamic properties
\begin{equation}
\tilde{p}\equiv \frac{\hbar p}{k_B T\sqrt{2m k_B T}},\quad \tilde{n}=\frac{\hbar n}{\sqrt{2m k_B T}},\quad \tilde{s}\equiv \frac{\hbar s}{k_B\sqrt{2m k_B T}}.
\end{equation} 
Consequently,  the dimensionless Yang-Yang TBAE  is given by
\begin{equation}\label{dimlessTBA}
\tilde{\varepsilon}(\tilde{k})=\tilde{k}^2-\tilde{\mu}-\frac{1}{2\pi}\int_{-\infty}^{\infty}\frac{2\tilde{c}}{\tilde{c}^2+(\tilde{k}-\tilde{q})^2}\ln\left(1+e^{-\tilde{\varepsilon}(\tilde{q})} \right)d \tilde{q}.
\end{equation} 
This serves as an equation of states for a whole temperature regime. 
This  integral equation can be numerically solved  by iteration method. 
In order to make a discretization in the variable space $\tilde{k}$, we need to find a proper cutoff $\tilde{k}_c$. The cutoff $\tilde{k}_c$ is determined by choosing $\ln(1+e^{-\tilde{\varepsilon}(\tilde{k}_c)})<10^{-9}$ because this term decreases quickly with increasing $\tilde{k}$, as it is shown in the Fig.~\ref{fig:S1}. 
The number of discretization from $\tilde{k}=0$ to $\tilde{k}=\tilde{k}_c$ is $N_k=1000$. Once we get $\tilde{\varepsilon}(\tilde{k})$ the dimensionless pressure $\tilde{p}$ can be obtained by
\begin{equation}
\tilde{p}\approx\frac{1}{2\pi}\int_{-\tilde{k}_c}^{\tilde{k}_c}\ln\left(1+e^{-\tilde{\varepsilon}(\tilde{k})} \right)d \tilde{k}.
\end{equation} 
Furthermore, we compare the pressure by taking different $N_k$ and find the difference can be negligible, as shown in the Fig.\ref{fig:S2}.

Based on the above numerical method, the density $\tilde{n}$ and entropy $\tilde{s}$ are obtained  by the derivatives of pressure \cite{Jiang15cpb}
\begin{eqnarray}
\tilde{n}&\approx-&\frac{1}{2\pi}\int_{-\tilde{k}_c}^{\tilde{k}_c}\frac{1}{1+e^{\tilde{\varepsilon}(\tilde{k})}}\varepsilon_{\mu} d \tilde{k},\\
\tilde{s}&\approx&\tilde{p}-\frac{1}{2\pi}\int_{-\tilde{k}_c}^{\tilde{k}_c}\frac{1}{1+e^{\tilde{\varepsilon}(\tilde{k})}}\left(\varepsilon_T-\tilde{\varepsilon}(\tilde{q})\right)d k,
\end{eqnarray} 
where the derivatives of $\varepsilon_{\mu}\equiv\frac{\partial \varepsilon(k)}{\partial \mu},\varepsilon_T\equiv\frac{\partial \varepsilon(k)}{\partial T}$ are given by
\begin{eqnarray}
\varepsilon_{\mu}&\approx &-1+\frac{1}{2\pi}\int_{-\tilde{k}_c}^{\tilde{k}_c}\frac{2\tilde{c}}{\tilde{c}^2+(\tilde{k}-\tilde{q})^2}\frac{1}{e^{\tilde{\varepsilon}(\tilde{q})}+1}\varepsilon_{\mu} d \tilde{q},\label{eq:deph}\\
\varepsilon_T&\approx & \tilde{\varepsilon}(\tilde{k})-\tilde{k}^2+\tilde{\mu}+\frac{1}{2\pi}\int_{-\tilde{k}_c}^{\tilde{k}_c}\frac{2\tilde{c}}{\tilde{c}^2+(\tilde{k}-\tilde{q})^2}\frac{1}{e^{\tilde{\varepsilon}(\tilde{q})}+1}\left(\varepsilon_T-\tilde{\varepsilon}(\tilde{q})\right) d \tilde{q}.\label{eq:dephT}
\end{eqnarray} 
These two equations may   be numerically solved by iteration.

As an example, when $\tilde{c}$ approaches $+\infty$, the last integral term in the dimensionless TBA, Eq.~\ref{dimlessTBA}, can be ignored and the gas behaves like free fermions~\cite{YangYang69JMP}. In particular, at the critical point $\tilde{\mu} = 0$, the entropy per particle can be computed analytically:
\begin{eqnarray}
\frac{S_{\mathrm{c}}}{N} & = & \frac{3}{2} \frac{\mathrm{Li}{}_{3/2} (-1)}{\mathrm{Li}{}_{1/2} (-1)} \approx 1.89738,
\end{eqnarray}
where $\mathrm{Li}_{s}$ is a polylogarithmic function of order $s$. 

\section{Fractional Exclusion Statistics}
In 1991, Haldane \cite{Haldane91PRL} formulated a  description of the fractional exclusion statistics (FES)   based on a generalized Pauli exclusion principle. 
This  FES was further formulated by Wu \cite{Wu94PRL,Wu94note} and others~\cite{Isakov94PRL, Ha94prl}.  It has been proved that the 1D $\delta$-function interacting Bose gas can be mapped onto  ideal particles  with FES, see \cite{Wu94PRL,Wu94note,Guan07laserPL}. In this sense, the dynamical and statistical interactions are transmutable. This allows one to deal with the thermodynamic properties through Haldane's FES. 
In general, the relation between the interaction strength and FES parameter is very complicated. However, under a strong interaction strength, the system may be equivalent to an  ideal gas with a non-mutual FES, i.e. the FES parameter does not depends on the momenta of  particles. 
For such a non-mutual FES with a parameter $g$, the occupation number $f$ in a state with energy $\epsilon=\frac{\hbar^2\bm{k}^2}{2m}$ is given by
\begin{equation}
f(\epsilon)=\frac{1}{w+g},
\end{equation} where $w$ obeys
\begin{equation}\label{eq:w}
w^g(1+w)^{1-g}=e^{\frac{\epsilon-\mu}{k_B T}}=e^{\tilde{\bm{k}}^2-\tilde{\mu}}.
\end{equation} 
The thermodynamic properties in $D$ dimension, such as pressure $p$, energy density $E$, and particle density $n$ are given by
\begin{eqnarray}
p&=&\frac{k_B T}{(2\pi)^D}\int_{-\infty}^{\infty} \ln\frac{1+w}{w}d^D \bm{k},\\
E&=&\frac{1}{(2\pi)^D}\int_{-\infty}^{\infty} \frac{\epsilon}{w+g}d^D \bm{k},\\
n&=&\frac{1}{(2\pi)^D}\int_{-\infty}^{\infty} \frac{1}{w+g}d^D \bm{k},
\end{eqnarray}
and the entropy density $s$ can be obtained from thermodynamic relation $E=-p+\mu n+s T$. 

The left hand side of Eq.~\ref{eq:w} is monotonically  increasing with $w$. For a given $g$, $\tilde{\mu}$ and $\bm{\tilde{k}}$, Eq.(\ref{eq:w}) can be numerically solved by bisection method. The relative error of $w$ in our numerical calculation is less than $10^{-6}$. The cutoff $\tilde{k}_c$ and discrete number $N_k$ here are the same with the one for solving Yang-Yang equation.

\section{Comparison between Yang-Yang equation and FES}
\begin{figure}
	\centering
	\includegraphics[scale=0.6]{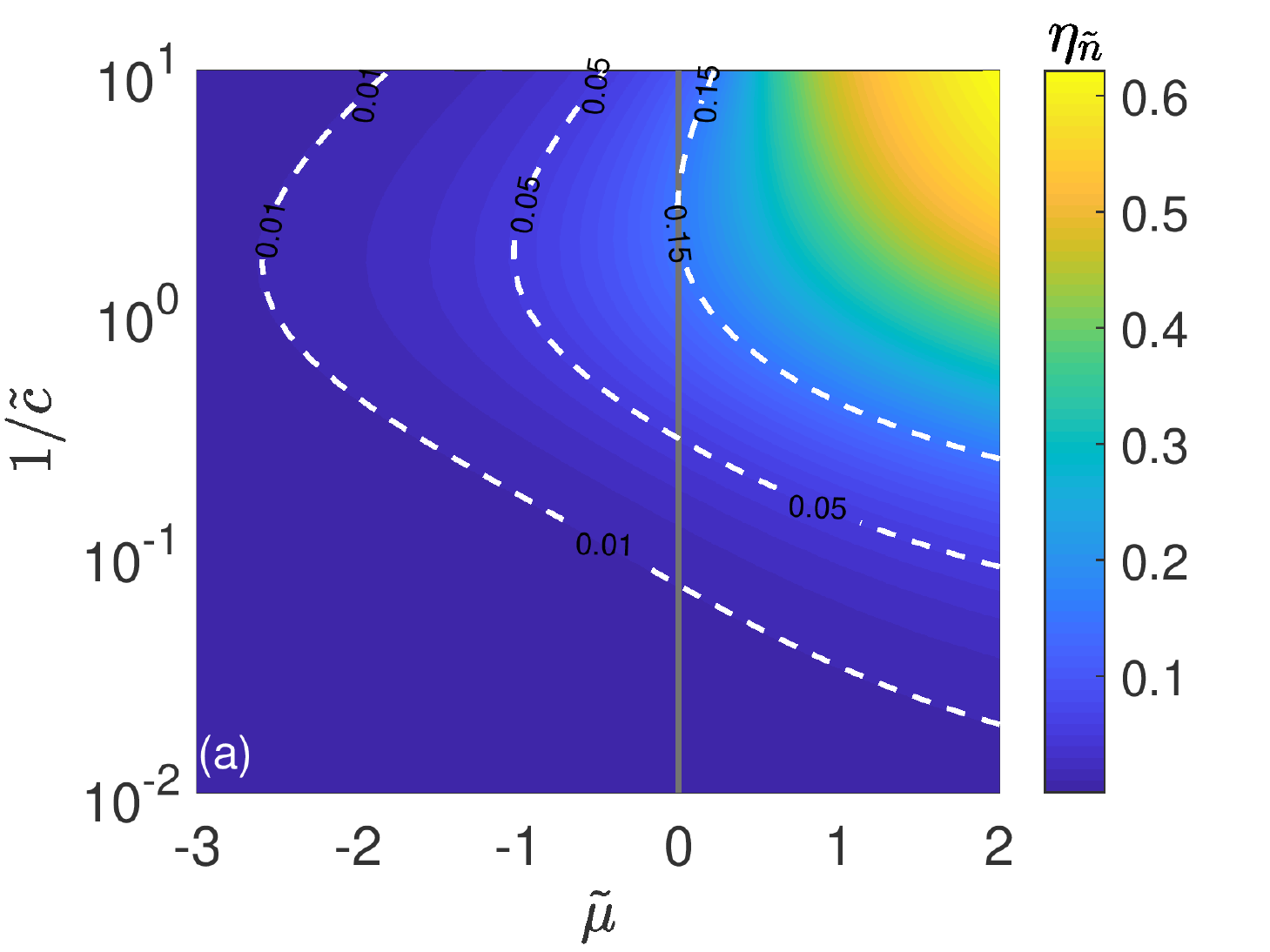}
	\includegraphics[scale=0.6]{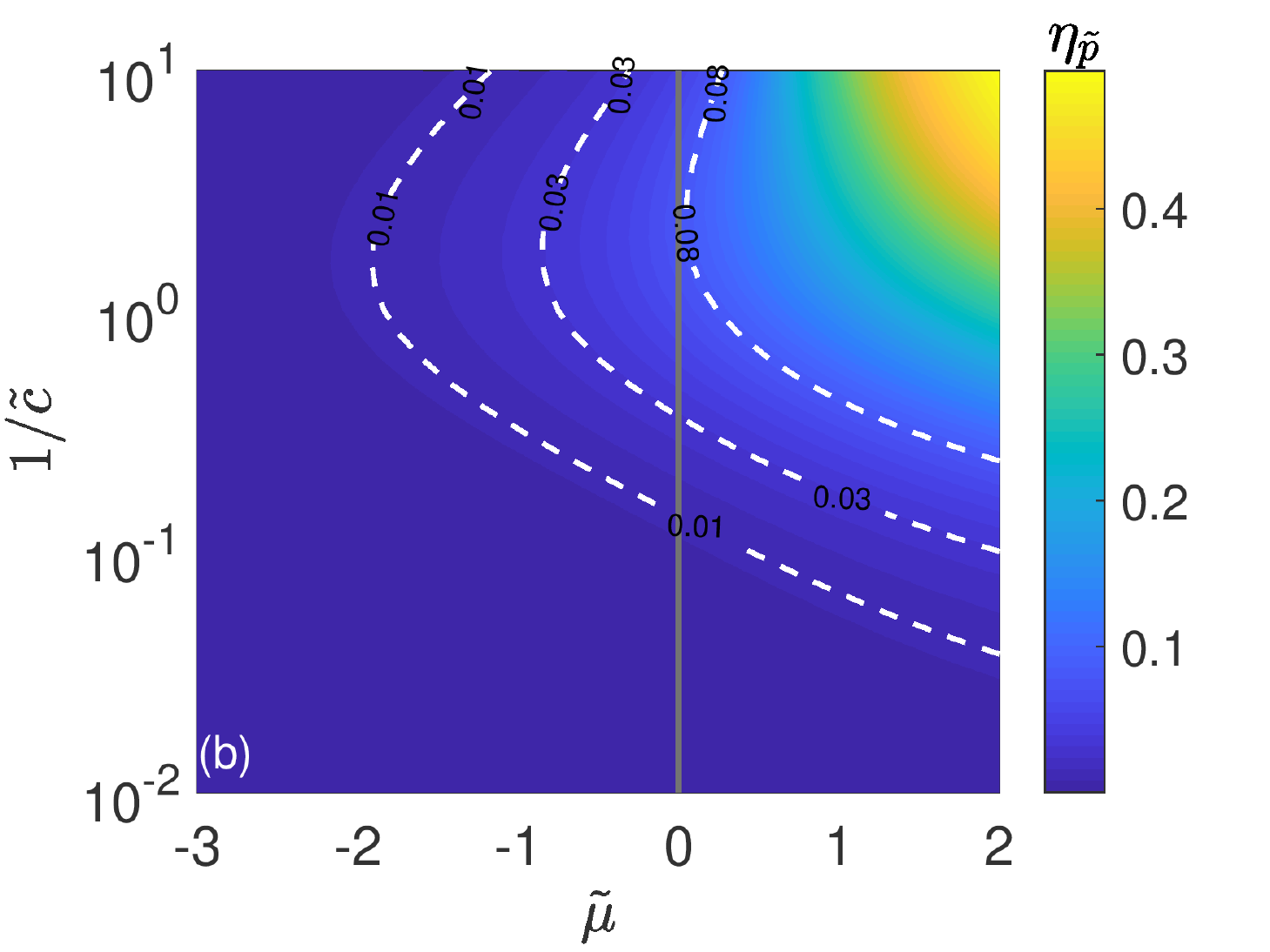}
	\includegraphics[scale=0.6]{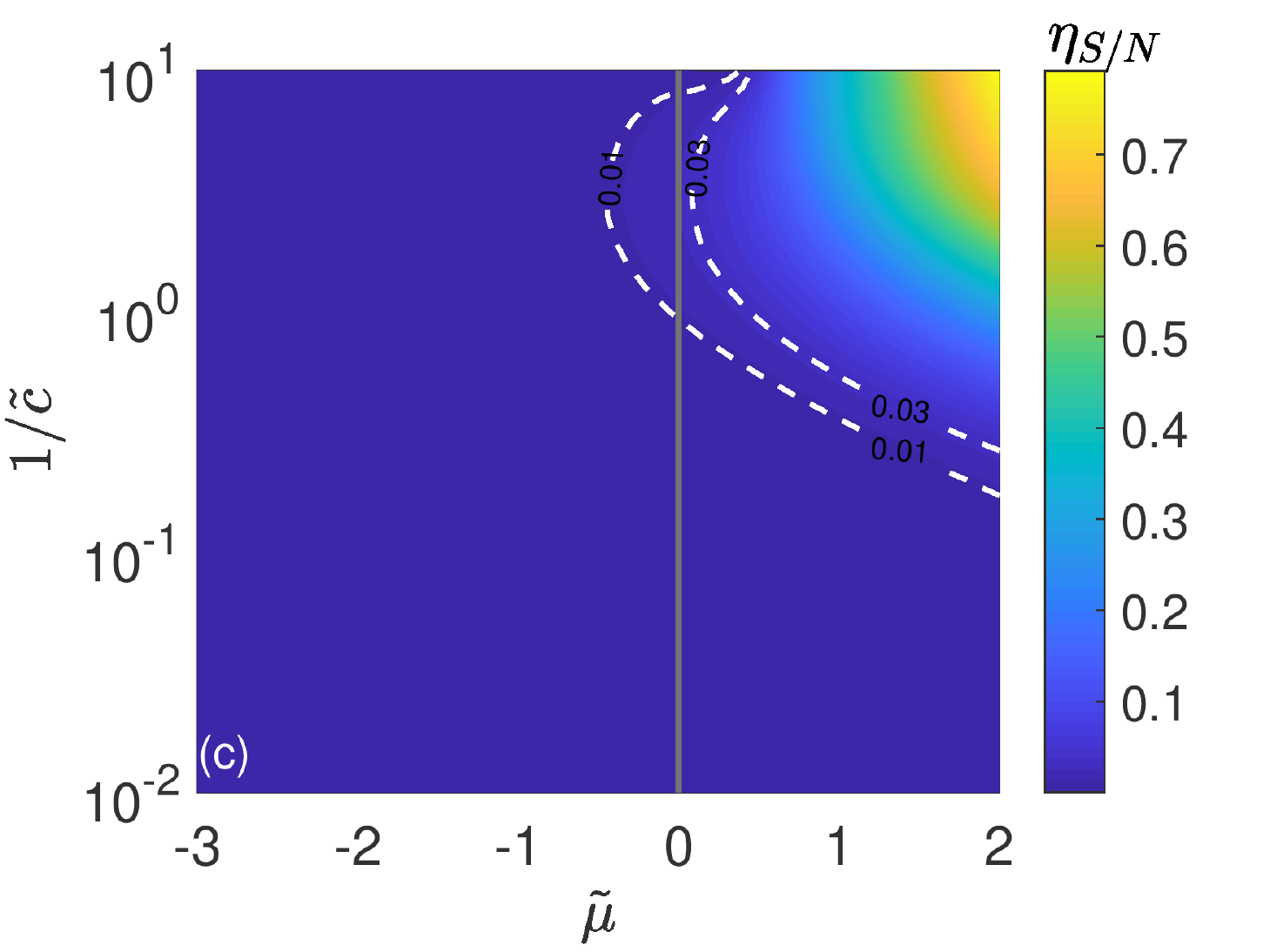}
	\caption{The comparison between interacting Bose gases and particles with non-mutual FES is shown in the chemical potential-interaction plane.    {\bf (a):} Contour plot  of the density's deviation $\eta_{\tilde{n}}$. {\bf (b):} Contour plot of the pressure's deviation $\eta_{\tilde{p}}$. {\bf (c):} Contour plot of the entropy per particle's deviation $\eta_{S/N}$. 
	}
	\label{fig:D}
\end{figure}

Although the mapping to non-mutual FES are obtained for critical point $\mu_c=0$, we can further extend it to $\tilde{\mu} = \frac{\mu - \mu_c}{T} \neq 0$. To quantify the agreement between solutions to the Yang-Yang equation and the computation results based on FES, we define
\begin{equation}
\eta_{\sigma}\equiv \frac{\vert\sigma_{Y}-\sigma_{F}\vert}{\sigma_{Y}}
\end{equation}where $\sigma=\tilde{n},\tilde{p},S/N$ and the subscripts ``Y'' and ``F'' denote the property obtained by Yang-Yang equation and FES, respectively. Our numerical results are shown in the Fig.~\ref{fig:D}. The contour plots of the  deviations of density  and pressure are  respectively shown near the critical point, where  $15\%$,  $8\%$ and $3\%$ derivations are marked. \\

\noindent \textit{The power of simple, non-mutual FES under strong interactions:} 

For a strong interaction, i.e.  $\tilde{c}\gg1$, these thermodynamic properties obtained from the  interacting system and  from the ideal particles with FES are in excellent agreement, even  for large $\tilde{\mu}$. The bottom part of each panel in Fig.~\ref{fig:D} illustrates this point.\\

\noindent \textit{The need of mutual FES under weak interactions:}  

\begin{figure}
	\centering
	\includegraphics[scale=0.6]{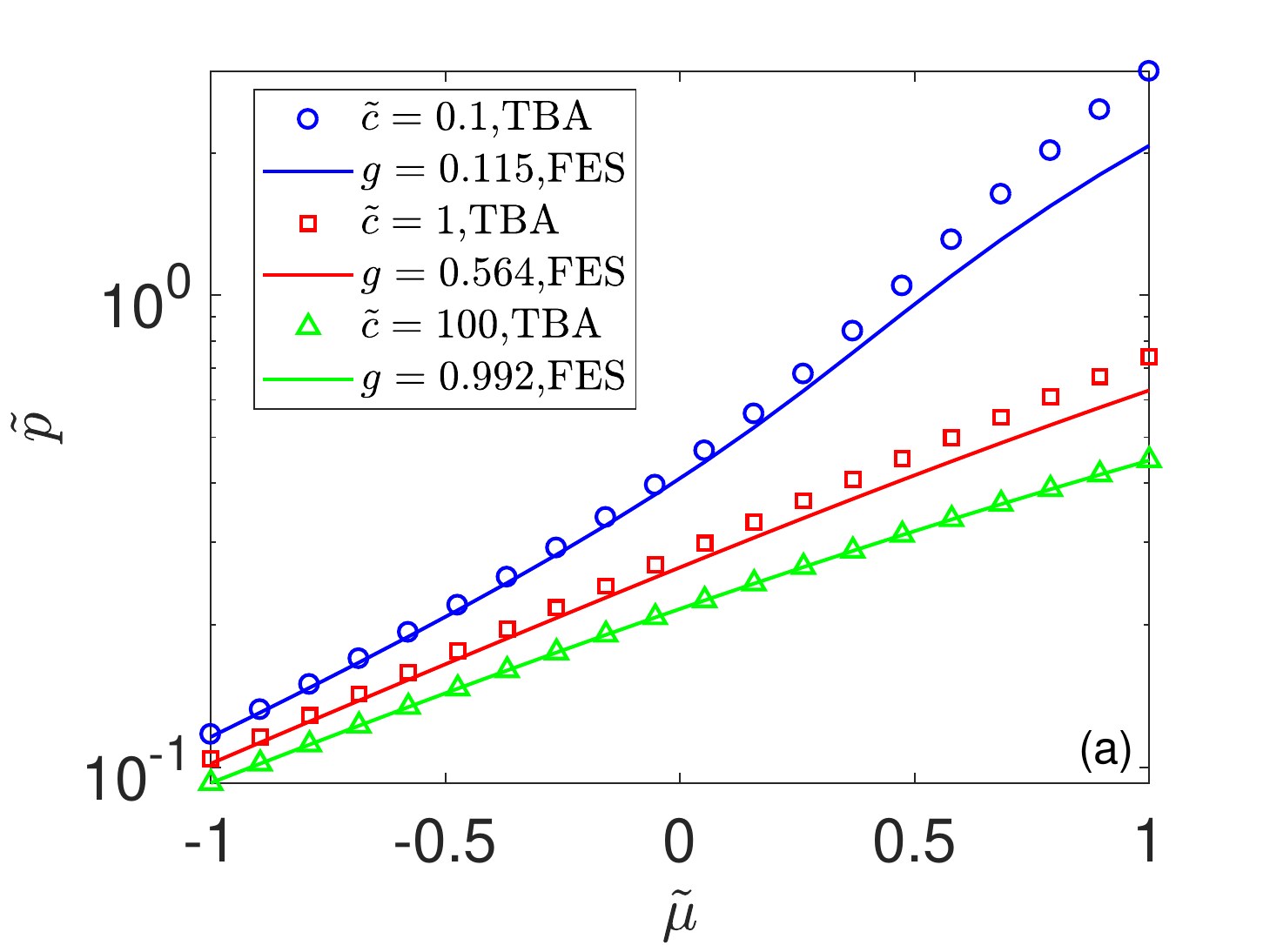}
	\includegraphics[scale=0.6]{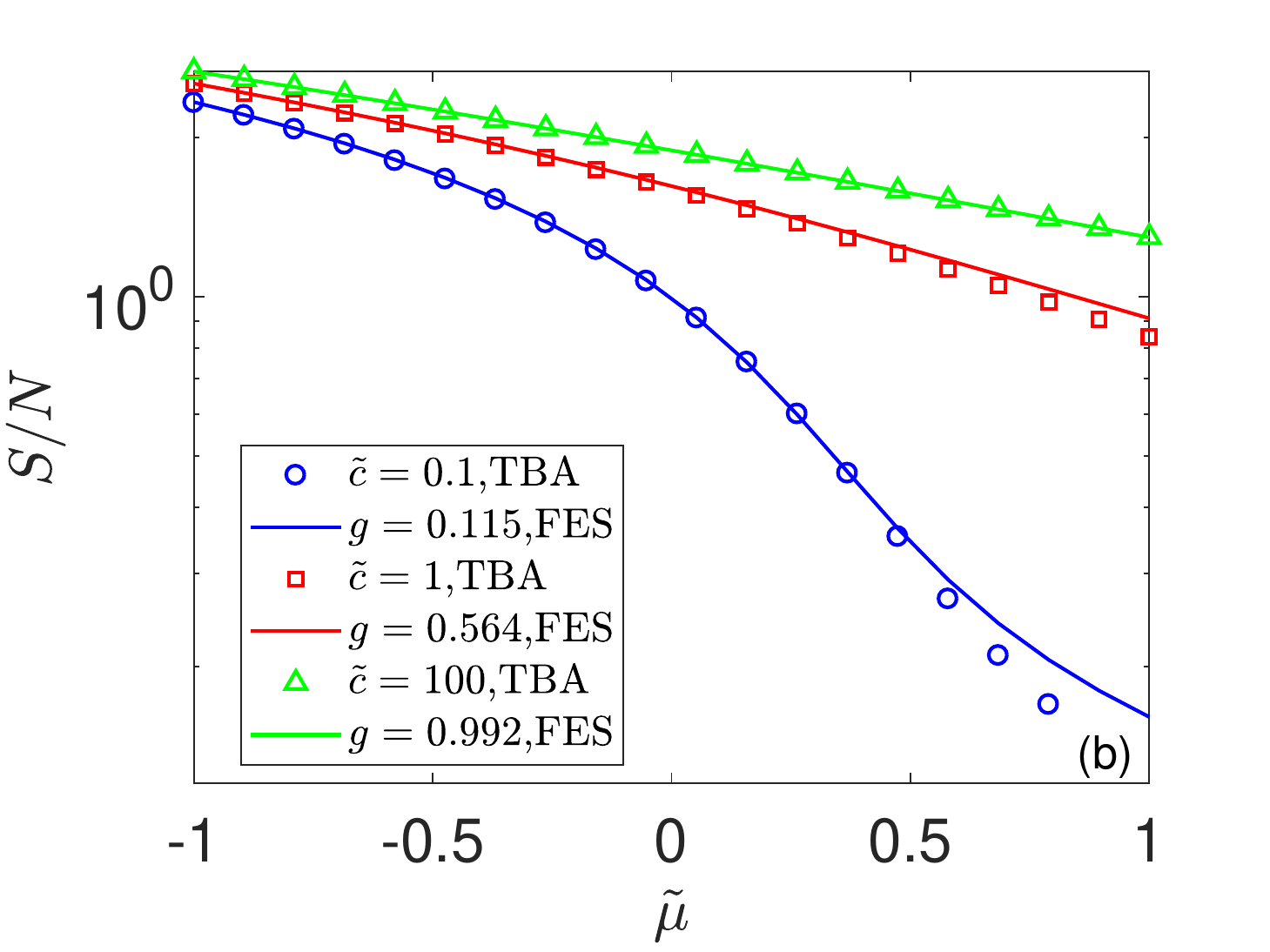}
	\caption{Scaled pressure $\tilde{p}$ and entropy per particle $S/N$ as a function of scaled chemical potential $\tilde{\mu}$: 1D interacting Bose gases (open symbols) compared to 1D non-interacting particles with FES (lines).  }
	\label{fig:p_and_sovn}
\end{figure}

On the other hand, under weak interactions, even when $S/N$ shows fairly good agreement (Fig.~\ref{fig:p_and_sovn}, $\tilde{c} = 0.1$, $-1 \le \tilde{\mu} < 0.5$), $\tilde{n}$ and $\tilde{p}$ show noticeable discrepancies in the same range, in particular for positive $\tilde{\mu}$ values, see   Fig.~4(a) in the main text and Fig.~\ref{fig:p_and_sovn} here, respectively. Thus these discrepancies are not caused by inaccuracy of our mapping formulae (Eqs.~7 and 6 in the main text) and cannot be alleviated by choosing a different ``effective FES parameter $g$''. Rather, such discrepancies in $\tilde{n}$ and $\tilde{p}$ are primarily due to the need of including more complex mutual FES effects~\cite{Wu94note}.

\hspace{1cm}


\section{Dimension analysis and dimensionless quantities}
In this section and the following two sections, we will explicitly explain how we derive the dimensionless observables (particle density, pressure and entropy per particle, especially at the quantum critical point) in  Bose gas in the continuous space based on the simulations for the Bose-Hubbard model on the lattice.

\hspace{1cm}

\hspace{1cm}

For $D$-dimensional ultracold Bose gas, the Hamiltonian could be written as~\cite{Greiner03thesis, AMR04thesis}
\begin{eqnarray}
\mathcal{H}=\sum_{i=1}^N\left(-\vec{\nabla}_i^2-\mu\right)+c\sum_{i\ne j}\delta(\mathbf{r}_i-\mathbf{r}_j).
\end{eqnarray}
This Hamiltonian has an equivalent field theory form
\begin{eqnarray}
\mathcal{H}=\int d^D r\left\{\hat{\psi}^\dagger(\mathbf{r})\left(-\nabla^2-\mu\right)\hat{\psi}(\mathbf{r})+c\hat{\psi}^\dagger(\mathbf{r})\hat{\psi}^{\dagger}(\mathbf{r})\hat{\psi}^{}(\mathbf{r})\hat{\psi}^{}(\mathbf{r})\right\}, \label{eq:H_g}
\end{eqnarray}
where $\hat{\psi}(\mathbf{r})$ is the wave function operator at the position $\mathbf{r}$.

By doing dimension analysis to Eq.~\ref{eq:H_g}, we obtain the dimension of the wavefunction operator and the parameters, that is
\begin{eqnarray}
&&[\hat{\psi}(\mathbf{r})]=L^{-D/2}, \\
&&[c]=EL^D, \\
&&[\mu]=E,
\end{eqnarray}
where $E$ and $L$ represent the dimension of energy and length, respectively.

If we choose $E$ to be the unit of energy and $\lambda$ to be the unit of length, we have the following relation equations between the quantities and their corresponding dimensionless quantities
\begin{eqnarray}
\mathbf{r}&=&\lambda \tilde{\mathbf{r}} \label{eq:x_tildex}\\ 
L&=&\lambda \tilde{L}\\
\nabla&=&\lambda^{-1}\tilde{\nabla}\\
\mu&=&E\tilde{\mu}\\
c&=&E\lambda^D\tilde{c}\\
\hat{\psi}(\mathbf{r})&=&\lambda^{-D/2}\tilde{\psi}(\tilde{\mathbf{r}}).
\end{eqnarray}
Here, we add a tilde to each of the symbols to denote their dimensionless quantities; for convenience, the dimensionless quantity for $\hat{\psi}$ is denoted as $\tilde{\psi}$ with the hat dropped.

From Eq. \ref{eq:x_tildex}, the Jacobian determinant is
\begin{eqnarray}
Det\left(\partial\mathbf{r}/\partial\tilde{\mathbf{r}}\right)=\lambda^D.
\end{eqnarray}
So, the integrals have the following relation equation
\begin{eqnarray}
\int d^D r F(\mathbf{r})=\left(\prod_{i=1}^D\int_0^{\tilde{L}} d\tilde{r}_i\right)Det\left(\partial\mathbf{r}/\partial\tilde{\mathbf{r}}\right)F(\lambda\tilde{\mathbf{r}})=\int d^D\tilde{r}\lambda^D F(\lambda\tilde{\mathbf{r}}).
\end{eqnarray}
%
%
%
Substituting these relation equations into Eq.~\ref{eq:H_g}, we can finally obtain the dimensionless form of the Hamiltonian
\begin{eqnarray}
\tilde{\mathcal{H}}(\tilde{c},\tilde{\mu})&=&\mathcal{H}/E \nonumber\\
&=&\int d^D\tilde{r}\left\{\tilde{\psi}^\dagger(\tilde{\mathbf{r}})\left(-\frac{1}{E\lambda^{2}}\tilde{\nabla}^2-\tilde{\mu}\right)\tilde{\psi}(\tilde{\mathbf{r}})+\tilde{c}\tilde{\psi}^\dagger(\tilde{\mathbf{r}})\tilde{\psi}^\dagger(\tilde{\mathbf{r}})\tilde{\psi}(\tilde{\mathbf{r}})\tilde{\psi}(\tilde{\mathbf{r}})\right\}, \label{eq:tilde_Hg0}
\end{eqnarray}
with the dimensionless parameters $\tilde{c}=c\lambda^{-D}E^{-1}$ and $\tilde{\mu}=\mu E^{-1}$.

Furthermore, if we take $E=T$ and $\lambda=\lambda_{dB}/(2\sqrt{\pi})=1/\sqrt{T}$, where $\lambda_{dB}=2\sqrt{\pi/T}$ is the thermal de Broglie wavelength, we can get $E\lambda^2=1$, and the dimensionless Hamiltonian becomes
\begin{eqnarray}
\tilde{\mathcal{H}}(\tilde{c},\tilde{\mu})&=&\mathcal{H}/T \nonumber\\
&=&\int d^D\tilde{r}\left\{\tilde{\psi}^\dagger(\tilde{\mathbf{r}})\left(-\tilde{\nabla}^2-\tilde{\mu}\right)\tilde{\psi}(\tilde{\mathbf{r}})+\tilde{c}\tilde{\psi}^\dagger(\tilde{\mathbf{r}})\tilde{\psi}^\dagger(\tilde{\mathbf{r}})\tilde{\psi}(\tilde{\mathbf{r}})\tilde{\psi}(\tilde{\mathbf{r}})\right\}, \label{eq:tilde_Hg}
\end{eqnarray}
with $\tilde{c}=cT^{D/2-1}$ and $\tilde{\mu}=\mu T^{-1}$.
The corresponding dimensionless forms of those observables we are interested in, the particle number density, pressure and entropy per particle, become
\begin{eqnarray}
\tilde{n}&=&n\lambda^{D}=nT^{-D/2}, \\
\tilde{p}&=&p\lambda^DE^{-1}=pT^{-(D/2+1)}, \\
\tilde{S}/\tilde{N}&=&S/N.
\end{eqnarray}


As for the Bose-Hubbard model simulated in our numerical part of work, its Hamiltonian reads
\begin{eqnarray}
\mathcal{H}_{_{BH}}=-t\sum_{<\vec{x},\vec{x}'>}\left(\hat{b}_{\vec{x}}^\dagger\hat{b}_{\vec{x}'}^{}+\hat{b}_{\vec{x}'}^\dagger\hat{b}_{\vec{x}}^{}\right)+\frac{U}{2}\sum_{\vec{x}}\hat{n}_{\vec{x}}(\hat{n}_{\vec{x}}-1)-\mu_{_{BH}}\sum_{\vec{x}}\hat{n}_{\vec{x}},
\end{eqnarray}
where $\vec{x}$ denotes the lattice site vector, $\hat{b}_{\vec{x}}^{}$ ($\hat{b}_{\vec{x}}^{\dagger}$) is the annihilation (creation) operator for bosons on the site $\vec{x}$, $\hat{n}_{\vec{x}}=\hat{b}_{\vec{x}}^\dagger\hat{b}_{\vec{x}^{}}$ is the particle number operator, and $<\vec{x},\vec{x}'>$ indicates the summation runs over all the nearest neighbor sites. The parameter $t$ is the tunneling parameter, $U$ is the onsite interaction strength, and $\mu_{_{BH}}$ is the chemical potential. Following the same approach, we can also obtain the dimensionless Hamiltonian for this model,
\begin{eqnarray}
\tilde{\mathcal{H}}_{_{BH}}=\frac{\mathcal{H}_{_{BH}}}{T}=-\tilde{t}\sum_{<\vec{x},\vec{x}'>}\left(\tilde{b}_{\vec{x}}^\dagger\tilde{b}_{\vec{x}'}^{}+\tilde{b}_{\vec{x}'}^{\dagger}\tilde{b}_{\vec{x}}\right)+\frac{\tilde{U}}{2}\sum_{\vec{x}}\tilde{n}_{\vec{x}}(\tilde{n}_{\vec{x}}-1)-\tilde{\mu}_{_{BH}}\sum_{\vec{x}}\tilde{n}_{\vec{x}}, \label{eq:dimensionless_HBH}
\end{eqnarray}
where $\tilde{b}_{\vec{x}}$ is the dimensionless quantity for $\hat{b}_{\vec{x}}$, which is exactly $\hat{b}_{\vec{x}}$ itself since $\hat{b}_{\vec{x}}$ is already dimensionless, and $\tilde{t}=t/T$, $\tilde{U}=U/T$ and $\tilde{\mu}_{_{BH}}=\mu_{_{BH}}/T$ are the corresponding dimensionless quantities for each parameter, and the dimensionless forms of the observables are as follows
\begin{eqnarray}
\tilde{n}_{_{BH}}&=&n_{_{BH}}, \\
\tilde{p}_{_{BH}}&=&p_{_{BH}}/T, \\
\tilde{S}_{_{BH}}/\tilde{N}_{_{BH}}&=&S_{_{BH}}/N_{_{BH}}.
\end{eqnarray}

\section{Mapping between Bose gases and the discrete Bose-Hubbard model}


Since $\int d^D \tilde{r}\tilde{\psi}^\dagger(\tilde{\mathbf{r}})\tilde{\nabla}^2\tilde{\psi}(\tilde{\mathbf{r}})=-\int d^D \tilde{r} \nabla\tilde{\psi}^\dagger(\tilde{\mathbf{r}})\cdot\nabla\tilde{\psi}(\tilde{\mathbf{r}})$, the dimensionless form of the Hamiltonian for Bose gas, Eq.~\ref{eq:tilde_Hg}, can also be written as
\begin{eqnarray}
\tilde{\mathcal{H}}=\int d^D \tilde{r}\left\{\tilde{ \nabla}\tilde{\psi}^\dagger(\tilde{\mathbf{r}})\cdot\tilde{\nabla}\tilde{\psi}(\tilde{\mathbf{r}})+\tilde{c}\tilde{\psi}^\dagger(\tilde{\mathbf{r}})\tilde{\psi}^\dagger(\tilde{\mathbf{r}})\tilde{\psi}(\tilde{\mathbf{r}})\tilde{\psi}(\tilde{\mathbf{r}})-\tilde{\mu}\tilde{\psi}^\dagger(\tilde{\mathbf{r}})\tilde{\psi}(\tilde{\mathbf{r}})\right\}\label{eq:H_g2}
\end{eqnarray}
In order to investigate the mapping relation between Bose gas and Bose-Hubbard model, we discretize the space into $N^D$ small cells with the side length $\tilde{\Delta}=\tilde{L}/N$, where $\tilde{L}$ is the size of the system, making the integral in Eq.~\ref{eq:H_g2} into sums:
\begin{eqnarray}
\tilde{\mathcal{H}}=\sum_{x_1=1}^N\sum_{x_2=1}^N...\sum_{x_D=1}^N \tilde{\Delta}^D \left\{\tilde{\nabla}\tilde{\psi}^\dagger_{\vec{x}}\cdot\tilde{\nabla}\tilde{\psi}_{\vec{x}}^{}+\tilde{c}\tilde{\psi}^\dagger_{\vec{x}}\tilde{\psi}^\dagger_{\vec{x}}\tilde{\psi}_{\vec{x}}^{}\tilde{\psi}_{\vec{x}}^{}-\tilde{\mu}\tilde{\psi}_{\vec{x}}^\dagger\tilde{\psi}_{\vec{x}}^{}\right\},
\end{eqnarray}
where $\tilde{\psi}_{\vec{x}}\equiv\tilde{\psi}(\vec{x}\tilde{\Delta})$, and $\vec{x}=(x_1,x_2,...,x_D)$ is the index of the cells with the integer components $x_i$ ranging from $1$ to $N$. $\tilde{\nabla}\tilde{\psi}_{\vec{x}}^{}$ are numerical differences of $\tilde{\psi}_{\vec{x}}^{}$ defined as
\begin{eqnarray}
\tilde{\nabla}\tilde{\psi}_{\vec{x}}=\frac{1}{\tilde{\Delta}}\left(\tilde{\psi}_{\vec{x}+\vec{e}_1}-\tilde{\psi}_{\vec{x}}^{},\tilde{\psi}_{\vec{x}+\vec{e}_2}^{}-\tilde{\psi}_{\vec{x}}^{},...,\tilde{\psi}_{\vec{x}+\vec{e}_D}^{}-\tilde{\psi}_{\vec{x}}^{}\right).
\end{eqnarray}
Thus, the discretized Hamiltonian becomes
\begin{eqnarray}
\tilde{\mathcal{H}}&=&-\tilde{\Delta}^{D-2}\sum_{<\vec{x},\vec{x}'>}\left(\tilde{\psi}^\dagger_{\vec{x}}\tilde{\psi}_{\vec{x}'}^{}+\tilde{\psi}^\dagger_{\vec{x}'}\tilde{\psi}_{\vec{x}}^{}\right)+\tilde{c}\tilde{\Delta}^D\sum_{\vec{x}}\tilde{\psi}^\dagger_{\vec{x}}\tilde{\psi}^\dagger_{\vec{x}}\tilde{\psi}_{\vec{x}}^{}\tilde{\psi}_{\vec{x}}^{}\nonumber\\
&&-\left(\tilde{\mu}\tilde{\Delta}^D-2D\tilde{\Delta}^{D-2}\right)\sum_{\vec{x}}\tilde{\psi}^\dagger_{\vec{x}}\tilde{\psi}_{\vec{x}}^{}
\end{eqnarray}
If we make the replacement
\begin{eqnarray}
\tilde{\psi}_{\vec{x}}^{}&=&\tilde{\Delta}^{-D/2}\tilde{b}_{\vec{x}},\nonumber\\
\tilde{\psi}^\dagger_{\vec{x}}&=&\tilde{\Delta}^{-D/2}\tilde{b}_{\vec{x}}^\dagger,
\end{eqnarray}
we can get
\begin{eqnarray}
\tilde{\mathcal{H}}
&=&-\tilde{\Delta}^{-2}\sum_{<\vec{x},\vec{x}'>}\left(\tilde{b}_{\vec{x}}^\dagger\tilde{b}_{\vec{x}'}^{}+\tilde{b}_{\vec{x}'}^\dagger\tilde{b}_{\vec{x}}^{}\right)+\tilde{c}\tilde{\Delta}^{-D}\sum_{\vec{x}}\tilde{n}_{\vec{x}}(\tilde{n}_{\vec{x}}-1)\nonumber\\
&& -\left(\tilde{\mu} -2D\tilde{\Delta}^{-2}\right)\sum_{\vec{x}}\tilde{n}_{\vec{x}}. \label{eq:dimensionless_Hg}
\end{eqnarray}

By comparing Eqs.~\ref{eq:dimensionless_Hg} and \ref{eq:dimensionless_HBH},
we obtain the mapping relations between Bose gas in the continuous space and Bose-Hubbard model in the lattice as follows
\begin{eqnarray}
\tilde{t}&=&\tilde{\Delta}^{-2},\\
\tilde{U}&=&2\tilde{c}\tilde{\Delta}^{-D},\\
\tilde{\mu}_{_{BH}}&=&\tilde{\mu}-2D\tilde{\Delta}^{-2},
\end{eqnarray}
or reversely,
\begin{eqnarray}
\tilde{\Delta}&=&\tilde{t}^{-1/2}=(T/t)^{1/2} \label{eq:tildeDelta}, \\
\tilde{c}&=&\frac{1}{2}\tilde{U}\tilde{\Delta}^D=\frac{1}{2}(U/t)(T/t)^{D/2-1}, \label{eq:tildec} \\
\tilde{\mu}&=&\tilde{\mu}_{_{BH}}+2D\tilde{t}=(\mu_{_{BH}}/t+2D)(T/t)^{-1}. \label{eq:tildemu}
\end{eqnarray}
From the last equation, we can know that the critical point in Bose gas $\mu=0$ corresponds to $\mu_{_{BH}}=-2Dt$, that is, the critical point for the phase transition from a vacuum phase to a quantum liquid phase in Bose-Hubbard model.

With some more analysis, we can also obtain the mapping relationships between the observables in both models as follows
\begin{eqnarray}
\tilde{n}&=&\tilde{n}_{_{BH}}\tilde{\Delta}^{-D}=n_{_{BH}}(T/t)^{-D/2}, \label{eq:tilde_n} \\
\tilde{p}&=&\tilde{p}_{_{BH}}\tilde{\Delta}^{-D}=(p_{_{BH}}/t)(T/t)^{-(D/2+1)}, \label{eq:tilde_p} \\
\tilde{S}/\tilde{N}&=&\tilde{S}_{_{BH}}/\tilde{N}_{_{BH}}=S_{_{BH}}/N_{_{BH}}. \label{eq:tilde_SN}
\end{eqnarray}
What we need to notice is that the corresponding dimensionless observables in Bose-Hubbard model for $\tilde{n}$ and $\tilde{p}$ are not simply $\tilde{n}_{_{BH}}$ and $\tilde{p}_{_{BH}}$.

\section{Extrapolation towards zero temperature}

\subsection{The extrapolation protocol and its application to 1D Bose-Hubbard systems}
According to the discretization approximation above, we would expect that 
when the dimensionless parameters $\tilde{c}$, $\tilde{\mu}$ in Bose gas model and ratios $U/t$, $\mu_{_{BH}}/t$, $T/t$ in Bose-Hubbard model are associated by Eq.~\ref{eq:tildec} and Eq.~\ref{eq:tildemu}, the corresponding dimensionless observables in two systems are equivalent to each other except a correction brought by the finite dimensionless spacing $\tilde{\Delta}$, which could be expressed by the following equation
\begin{eqnarray}
\tilde{O}(\tilde{g},\tilde{\mu})&=&\tilde{O}_{_{BH}}(U/t,\mu_{_{BH}}/t,T/t)+\tilde{f}(\tilde{g},\tilde{\mu},\tilde{\Delta}), \label{eq:tildeO_tildeDelta}
\end{eqnarray}
where $\tilde{O}$ and $\tilde{O}_{_{BH}}$ are two corresponding dimensionless observables in interacting Bose gas and Bose-Hubbard model respectively, and $\tilde{f}$ is the dimensionless correction function decaying to zero when $\tilde{\Delta}$ is approaching to zero, that is
\begin{eqnarray}
\lim\limits_{\tilde{\Delta}\rightarrow 0}\tilde{f}(\tilde{g},\tilde{\mu},\tilde{\Delta})=0. \label{eq:limit_tildef_tildeDelta}
\end{eqnarray}
From Eq.~\ref{eq:tildeDelta}, we know that the correction function could be rewritten as a function of the temperature, that is $\tilde{f}(\tilde{g},\tilde{\mu},T/t)$, and $\tilde{\Delta}\rightarrow 0$ is equivalent to $T/t\rightarrow 0$. Thus Eq. \ref{eq:tildeO_tildeDelta} and Eq. \ref{eq:limit_tildef_tildeDelta} becomes
\begin{eqnarray}
\tilde{O}(\tilde{g},\tilde{\mu})=\tilde{O}_{_{BH}}(U/t,\mu_{_{BH}}/t,T/t)+\tilde{f}(\tilde{g},\tilde{\mu},T/t),
\end{eqnarray}
and
\begin{eqnarray}
\lim\limits_{T/t\rightarrow 0}\tilde{f}(\tilde{g},\tilde{\mu},T/t)=0.
\end{eqnarray}
So, for the Bose-Hubbard model, only when $T/t\rightarrow 0$ will the dimensionless observables collapse to the corresponding dimensionless ones in continuous-space Bose gases.

Concluding from the analysis above, in order to obtain a dimensionless observable in Bose gas at $\tilde{c}$ and $\tilde{\mu}$, we can first compute the corresponding dimensionless observables in Bose-Hubbard model at different temperatures with the parameter ratios $U/t$ and $\mu_{_{BH}}/t$ determined by Eq.~\ref{eq:tildec} and Eq.~\ref{eq:tildemu}, and then extrapolate the results towards zero temperature, which corresponds to the results in a Bose gas with no lattices. 


As a typical example shown in Fig. \ref{fig:ScN_T_1d_tildec1}, we measure and compute the dimensionless observables $\tilde{S}/\tilde{N}$ (equivalent with $S/N$) and $\tilde{ n}$ at the critical point in one-dimensional Bose-Hubbard model by the formulas Eq. \ref{eq:tilde_n} and \ref{eq:tilde_SN} with $\tilde{c}=1$ and the temperature $T/t$ varying from $1.0$ to $0.1$. We extrapolate these results to zero temperature by different formulas with different ranges of the temperature, and according to the distribution of these extrapolation results, we obtain the final estimates $S_c/N(\tilde{c}=1)=1.602(12)$ and $\tilde{n}_c(\tilde{c}=1)=0.28042(66)$, which are consistent with the theoretical results $1.602509$ and $0.280377$ obtained from solutions to the thermodynamic Bethe ansatz  equation. Here, during the extrapolation, we apply rigorous statistics standards and only accept those fitting routines that can describe all data in the fitting ranges within $3$ times of their error bars.

\subsection{A discussion on scale invariance and the extrapolation towards zero temperature}
As shown by Eq.~\ref{dimlessTBA}, the 1D interacting boson system studied in this work satisfies ``scale invariance'', namely,  numerical or experimental data taken at different temperatures can ``collapse'' onto universal functions ($S/N$ versus $\tilde{\mu}$, $\tilde{n}$ versus $\tilde{\mu}$, $\tilde{p}$ versus $\tilde{\mu}$) when the parameters (quasi-momentum $k$, interaction strength $c$) and thermodynamic quantities are scaled properly according to the thermal de Broglie wavelength. On the other hand, a lattice gas system does not strictly satisfy  scale invariance. To approach these universal functions using 1D Bose-Hubbard model, we need to reach a parameter regime where the dimensionless lattice spacing $\tilde{\Delta}$ is much smaller than all other relevant dimensionless length scales -- and thus decoupled from the physical properties of the system.  This is equivalent to requiring that the thermal de Broglie wavelength be much larger than all other relevant length scales. We further convert this requirement into a final criterion that the temperature be much lower than  all other relevant energy scales -- and thus decoupled from the physical properties of the system. Mathematically this is satisfied when $T/t$ becomes sufficiently small. This is the physical motivation of our protocol of ``extrapolation towards zero temperature''. In a practical simulation, all our numerical data are obtained under finite temperatures. So our extrapolation results represent the physical properties when the system temperature is much lower than  all other relevant energy scales. As long as the temperature does not play a role in influencing the physical properties of the system, we consider the purpose of the extrapolation protocol to be fulfilled. 



\begin{figure}[th]
	\centering
	\includegraphics[width=0.49\linewidth]{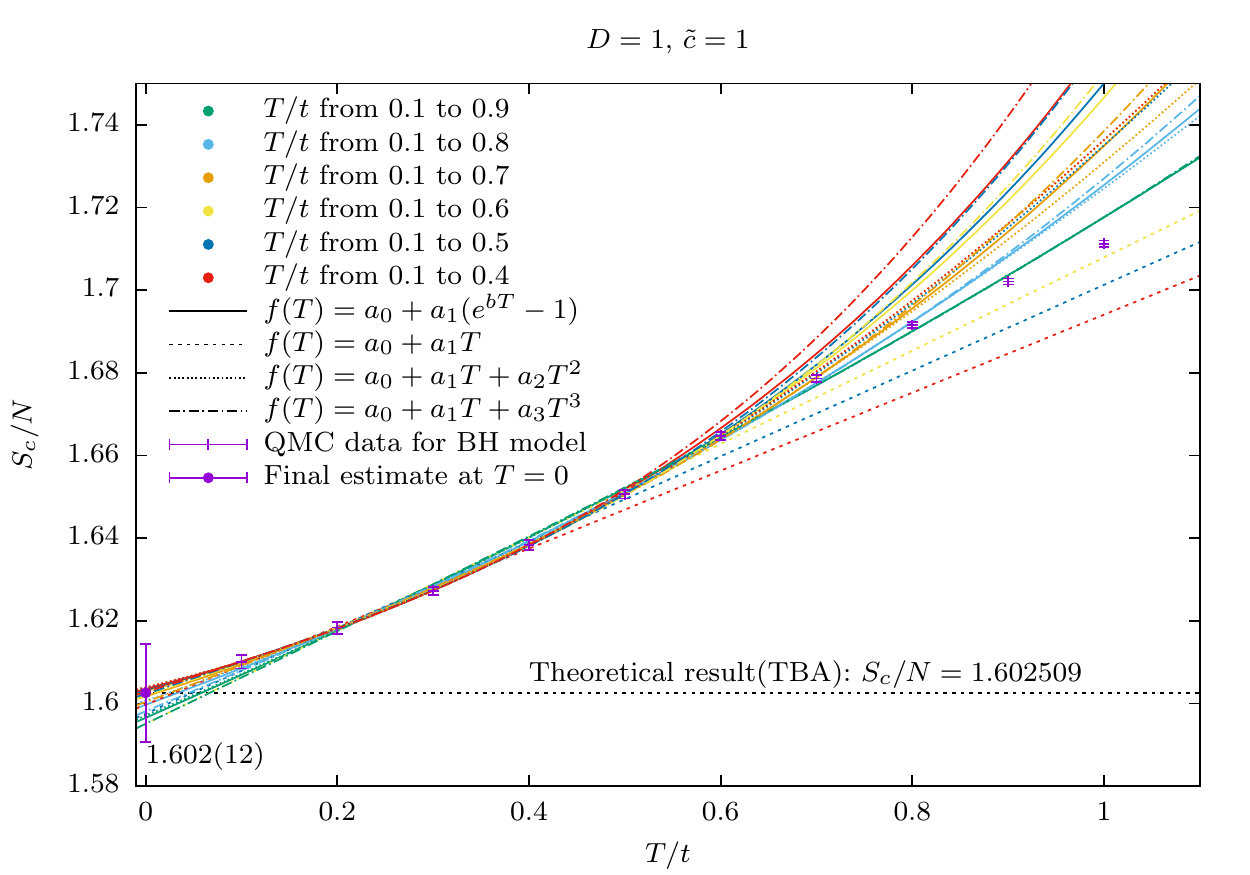}
	\includegraphics[width=0.49\linewidth]{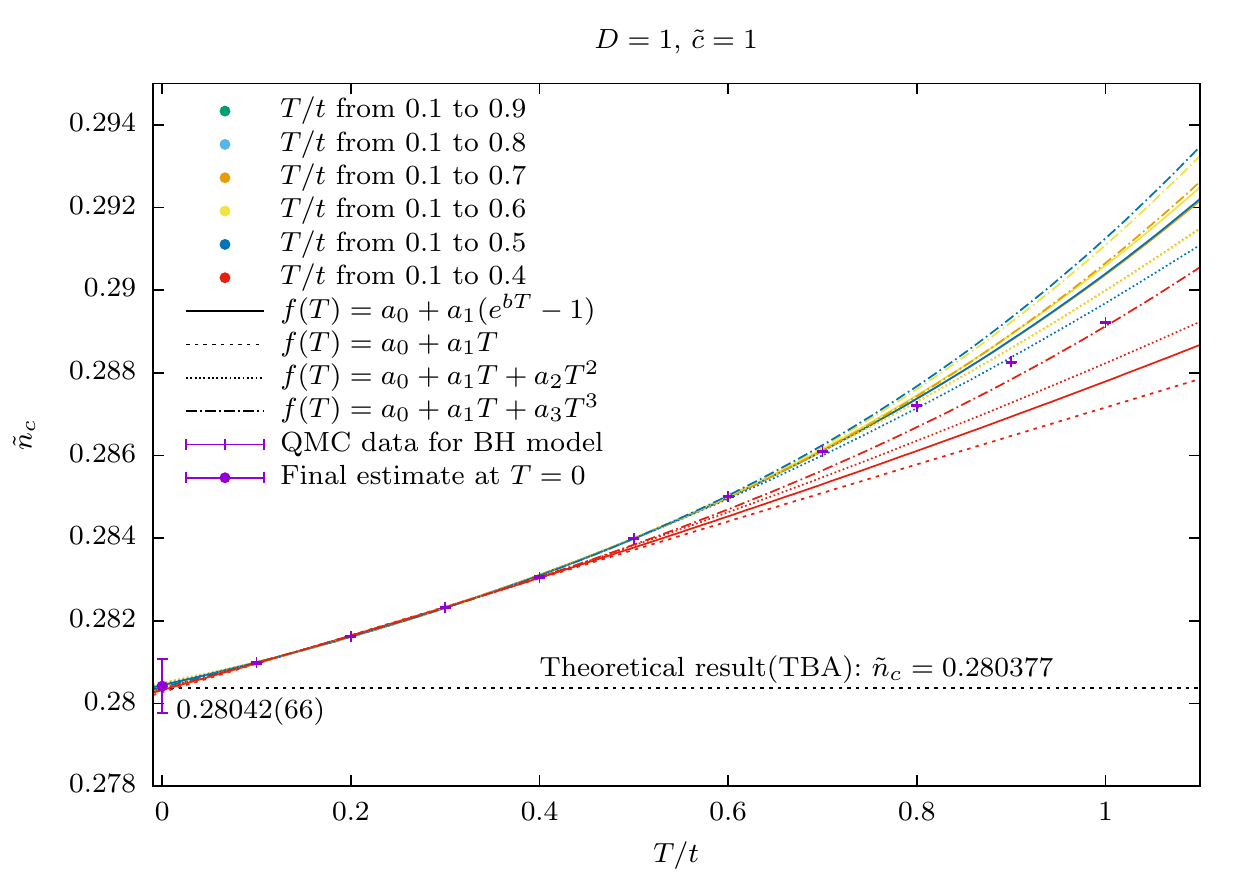}
	\caption{The entropy per particle and dimensionless particle number density at the quantum critical point, $S_c/N$ and $\tilde{n}_c$, vs.  $T/t$ at $\tilde{c}=1$ for one-dimensional Bose-Hubbard model. We apply different fitting formulas and fitting ranges of the temperature $T/t$ shown in the figure to extrapolate the results towards zero temperature. Different colors of the lines indicate different fitting ranges, while different dash types of the lines represent different fitting formulas. Besides the formulas shown in this figure, some higher order of polynomials are also applied to do the extrapolation. According to the distribution of these extrapolation results, we obtain the final estimates as $S_c/N(\tilde{c}=1)=1.602(12)$ and $\tilde{n}_c(\tilde{c}=1)=0.28042(66)$, which are consistent with the theoretical results $1.602509$ and $0.280377$ derived by solutions to the thermodynamic Bethe ansatz (TBA) equations.} \label{fig:ScN_T_1d_tildec1}
\end{figure}

In 1D, we know beforehand that the system satisfies scale invariance, such that the extrapolation towards zero temperature must have a well-defined limit. We indeed observe convergence in extrapolation (see Fig.~\ref{fig:ScN_T_1d_tildec1}), and observe that the QMC results (after extrapolation) agree excellently with the solutions to TBA equations (main text, Fig.~2). The statistical uncertainties of the extrapolated results reflect the accuracy of our QMC simulations and our confidence in using the QMC data to reveal the known physical and scaling properties of 1D Bose gases in continuous space. This computation serves as a calibration of our extrapolation protocol. When we apply this extrapolation protocol to 2D Bose-Hubbard lattice gases, where the corresponding continuous-space 2D Bose gases don't have an explicit equation like the TBA equation, \textit{ we do not presume a prerequisite that the continuous-space system satisfies scale invariance. Instead, we rely on the statistical uncertainties of the extrapolated results to provide understanding on the scaling behaviors and physical properties of the continuous-space Bose gases under sufficiently low temperatures.}

\begin{figure}
	\centering
	\includegraphics[scale=0.6]{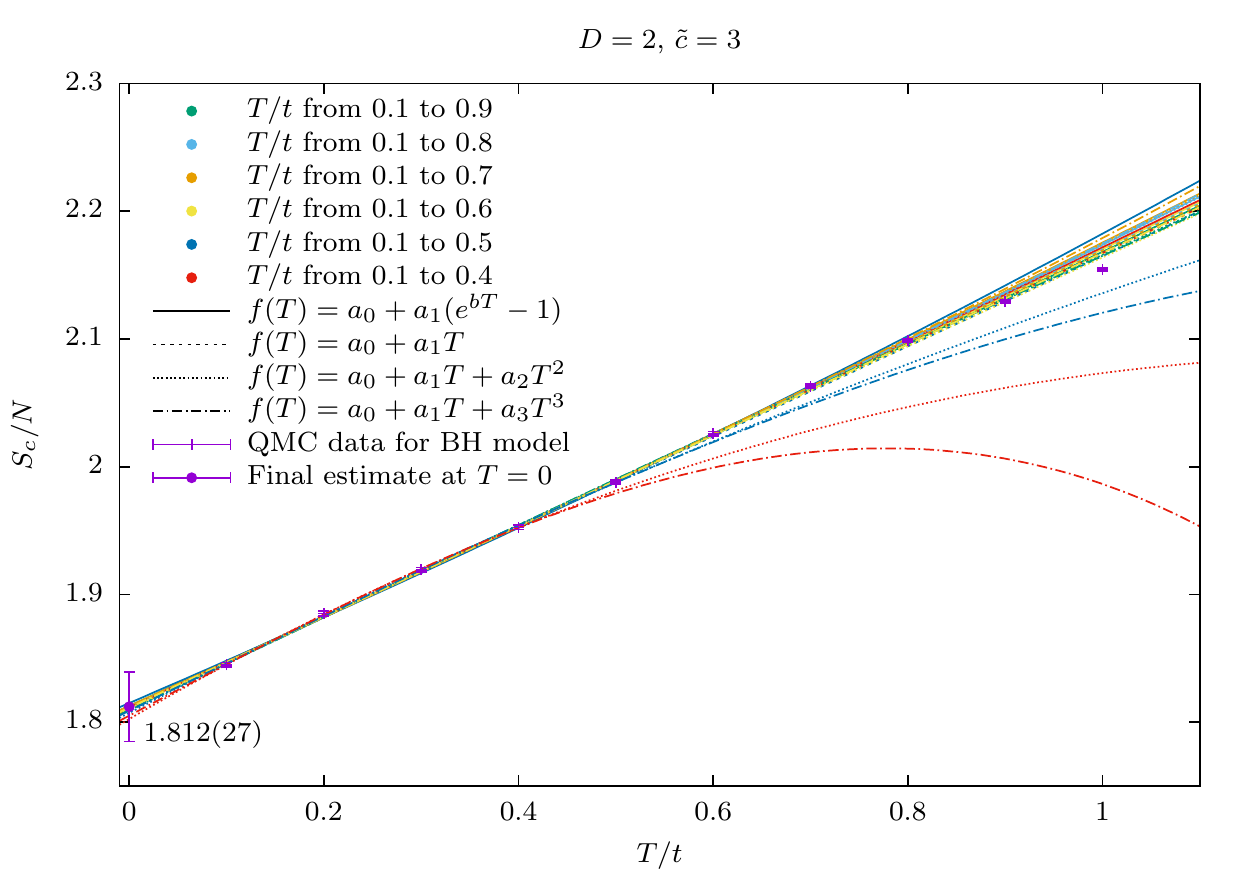}
	\includegraphics[scale=0.6]{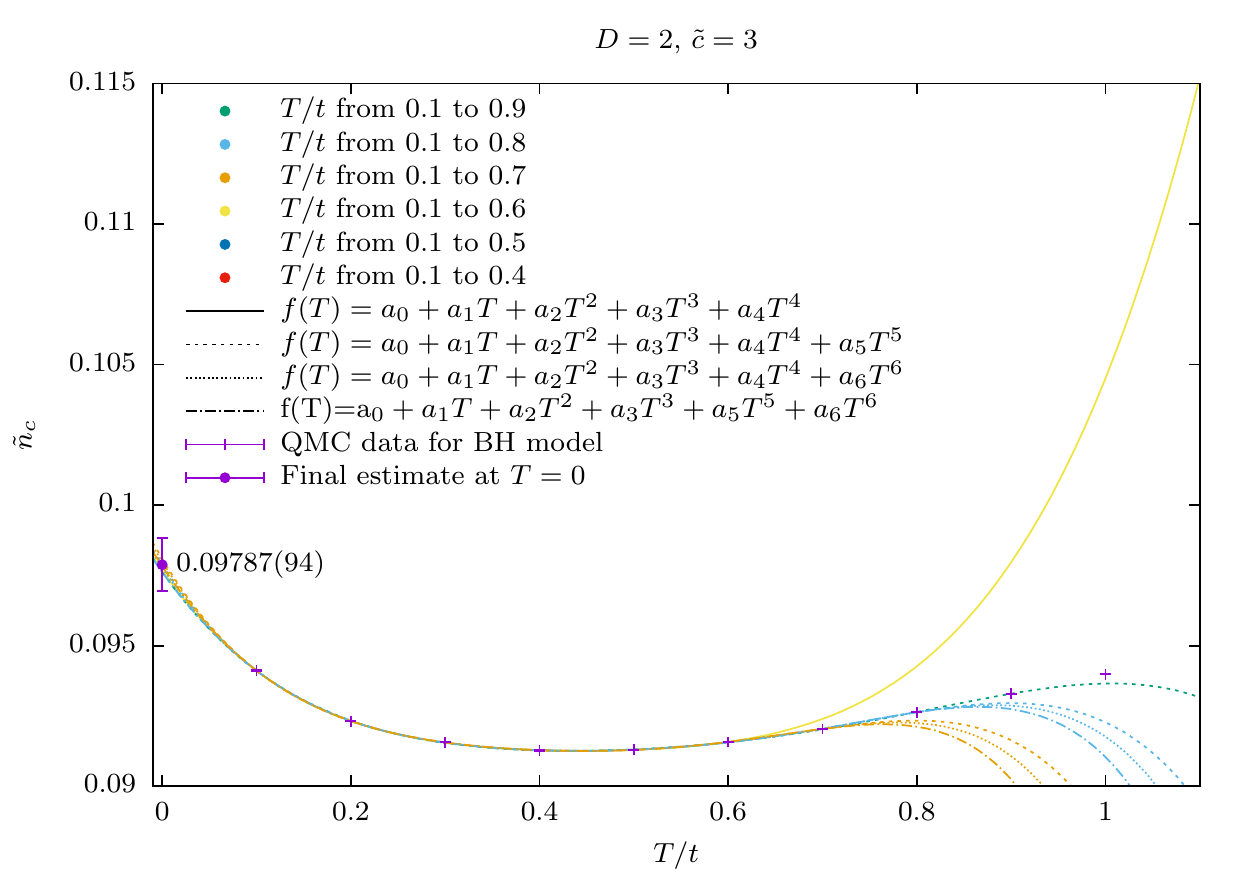}
	\includegraphics[scale=0.6]{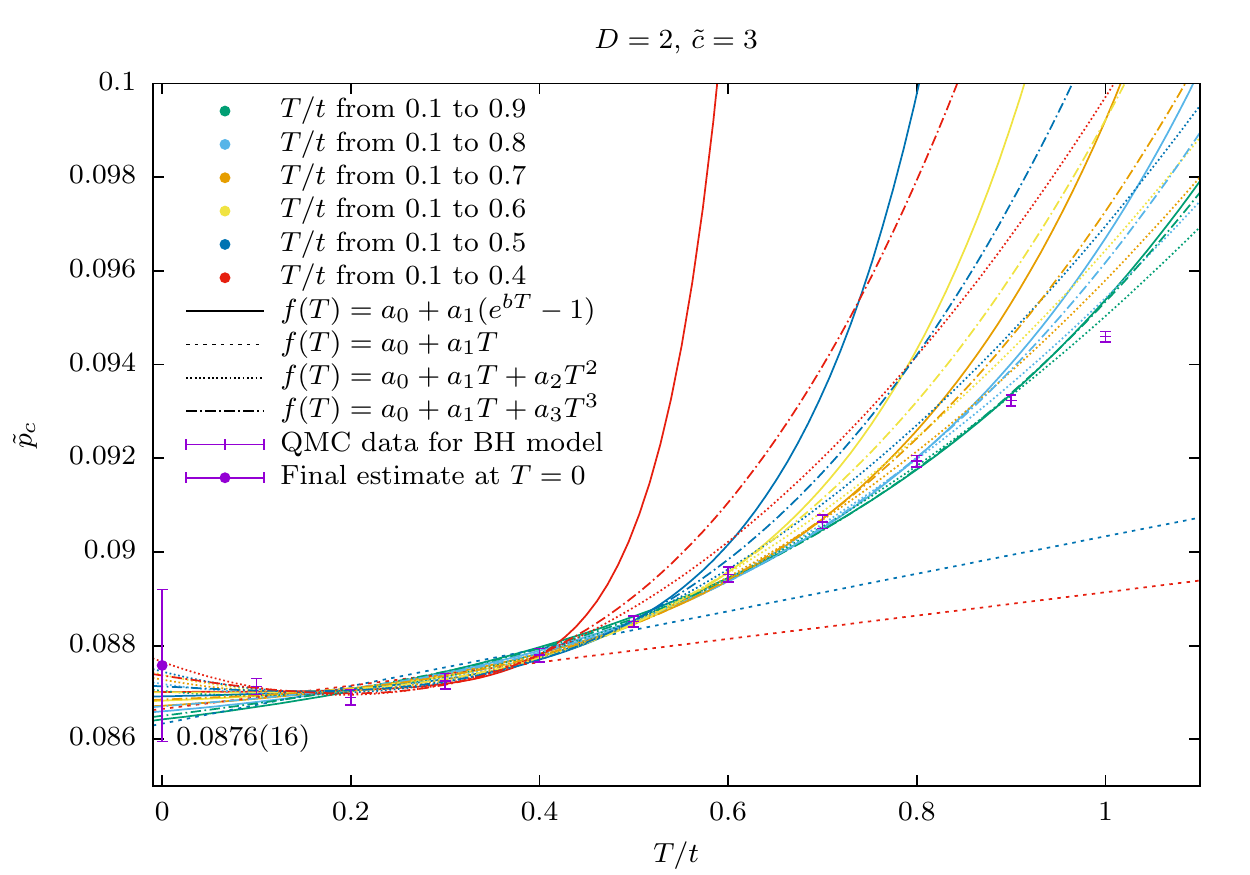}
	\caption{The entropy per particle,  dimensionless scaled particle number density, and scaled pressure at the quantum critical point: $S_c/N$, $\tilde{n}_c$, $\tilde{p}_c$ vs.  $T/t$ at $\tilde{c}_{\mathrm{2D}}=3$ for two-dimensional Bose-Hubbard model.  We apply different fitting formulas and fitting ranges of the temperature $T/t$ shown in the figure to extrapolate the results towards zero temperature. Different colors of the lines indicate different fitting ranges, while different dash types of the lines represent different fitting formulas.}
	\label{fig:2Dextrapolation}
\end{figure}

In an example shown in Fig.~\ref{fig:2Dextrapolation}, we study 2D Bose-Hubbard model at $\tilde{c}_{\mathrm{2D}} = U/(2t) = 3$ and determine $S_c/N$, $\tilde{n}_c$, $\tilde{p}_c$ with the temperature $T/t$ varying from 1.0 to 0.1. We extrapolate these results to zero temperature and obtain $S_c/N = 1.812(27)$, $\tilde{n}_c = 0.09787(94)$, and $\tilde{p}_c = 0.0876(16)$. We observe that the statistical uncertainties of the extrapolated results are fairly small, primarily because each individual numerical data point is  accurately determined and has a small error bar (within $1 \sim 3\%$ level). While these results can potentially be further improved by future simulations with even lower simulation temperatures (T/t being on the $10^{-2}$ to $10^{-3}$ level), the current extrapolation results are sufficient to reveal the physical properties ($S_c/N$, $\tilde{n}_c$, $\tilde{p}_c$) and scaling behaviors of the corresponding continuous-space Bose gases under sufficiently low temperatures.

As stated and shown in the main text (Figs.~3 and 4(b)), we obtain numerical results for 2D interacting Bose gases  using QMC simulation data and the above extrapolation protocol. Our results agree with a non-perturbative renormalization group (NPRG) computation~\cite{Rancon12PRA} for $\tilde{c}_{\mathrm{2D}} < 1$, and agree with experiments on 2D Bose gases without or with optical lattices~\cite{Zhang12Science,Dalibard11PRL,Hung11nature,Ha13PRL} for $ 0.05 \le \tilde{c}_{\mathrm{2D}} \le 4.2 $. These agreements confirm our estimate on the physical properties of  2D Bose gases based on extrapolation. Hence our results for $\tilde{c}_{\mathrm{2D}} = 100$, $200$, $1000$, and $\infty$, obtained using the same extrapolation method and with similarly small statistical uncertainties, further provide new insights into a system of continuous-space 2D Bose gases with strong repulsive interactions (and with no inelastic losses in the model) under sufficiently low temperatures.

%
%
%
\section{Measuring of the observables in Bose-Hubbard model}

In our work, we mainly focus on the following observables: particle density $n$, pressure $p$ and entropy per particle $S/N$. In this section, we will show how we derive these observables in Bose-Hubbard model.

\begin{figure}[th]
	\centering
	\includegraphics[width=0.8\linewidth]{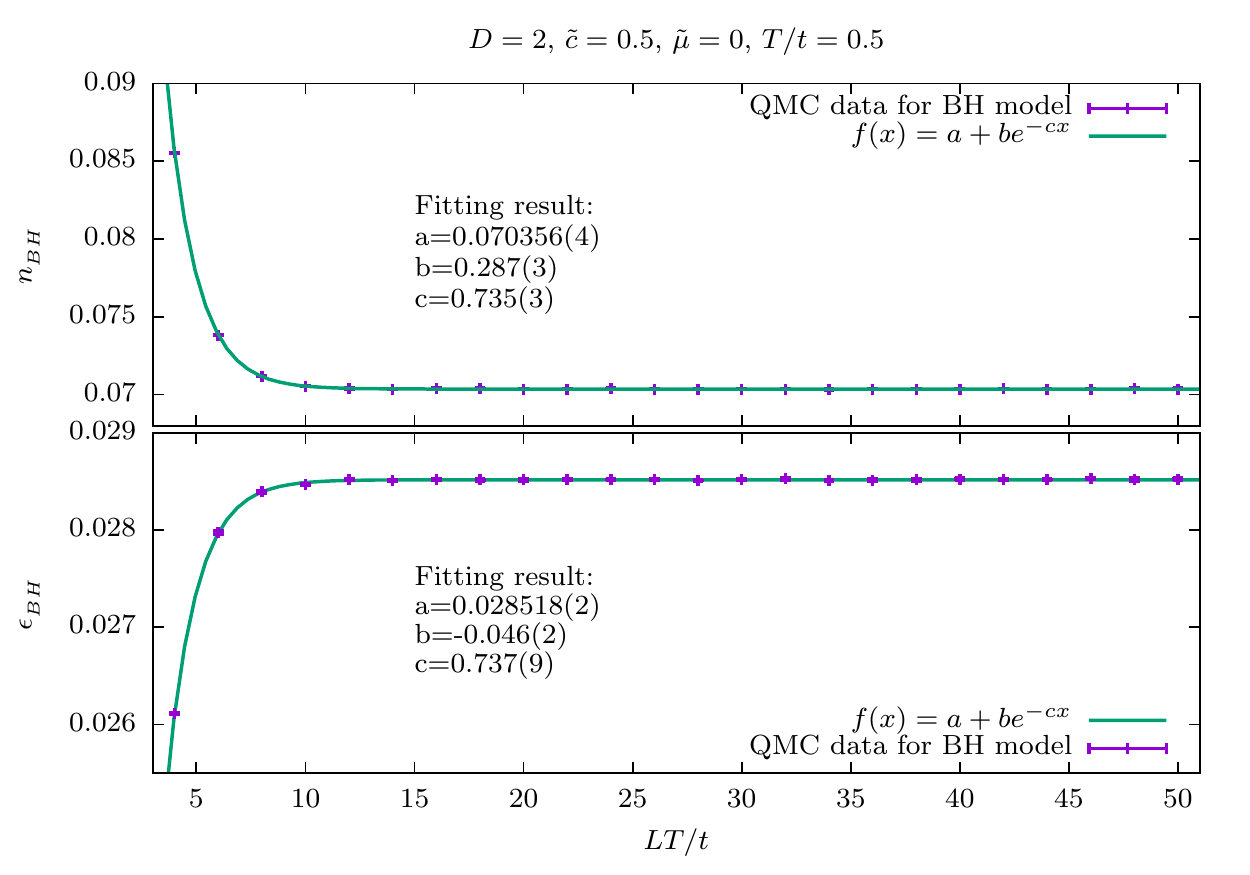}
	\caption{The particle number density and the grand energy density in Bose-Hubbard model, $n_{_{BH}}$ and $\epsilon_{_{BH}}$, vary with $LT/t$. Both curves could be well fitted by the formula $f(x)=a+be^{-cx}$, and according to the fitting results, even when $LT/t=20$, the relative errors brought by the finite size of the system are of the order $10^{-6}$, which is pretty small compared with the statistical relative errors (typically $10^{-4}\sim 10^{-3}$). } \label{fig:n_Egrand_2d_tildec0.5}
\end{figure}

For both one-dimensional and two-dimensional systems, we apply worm algorithm in path-integral representation to simulate the Bose-Hubbard model by quantum Monte Carlo (QMC) method \cite{qmc98pla,qmc98jetp}. During the simulations, we can directly measure the particle number density $n_{_{BH}}$ and the grand energy density $\epsilon_{_{BH}}\equiv\frac{1}{V}\langle \mathcal{H}_{_{BH}}\rangle$, where $V$ is the volume of the system, and in lattice model is just the total number of the sites. 
In order to ensure the data we obtained are in the thermodynamic limit, we keep the size of the system we simulated to be at least $L=40t/T$, except when $T/t=0.1$ we choose $L=20t/T$ for some of them. A typical example in two dimension is shown in Fig.~\ref{fig:n_Egrand_2d_tildec0.5}, and it presents that the system sizes we choose are large enough to allow us to neglect the errors brought by finite system sizes.


Since the particle number density for Bose-Hubbard model $n_{_{BH}}$ could be measured directly in our simulations, we will principally introduce how we derive the pressure $p_{_{BH}}$ and the entropy $S_{_{BH}}$ for Bose-Hubbard model.

\begin{figure}[th]
	\centering
	\includegraphics[width=0.8\linewidth]{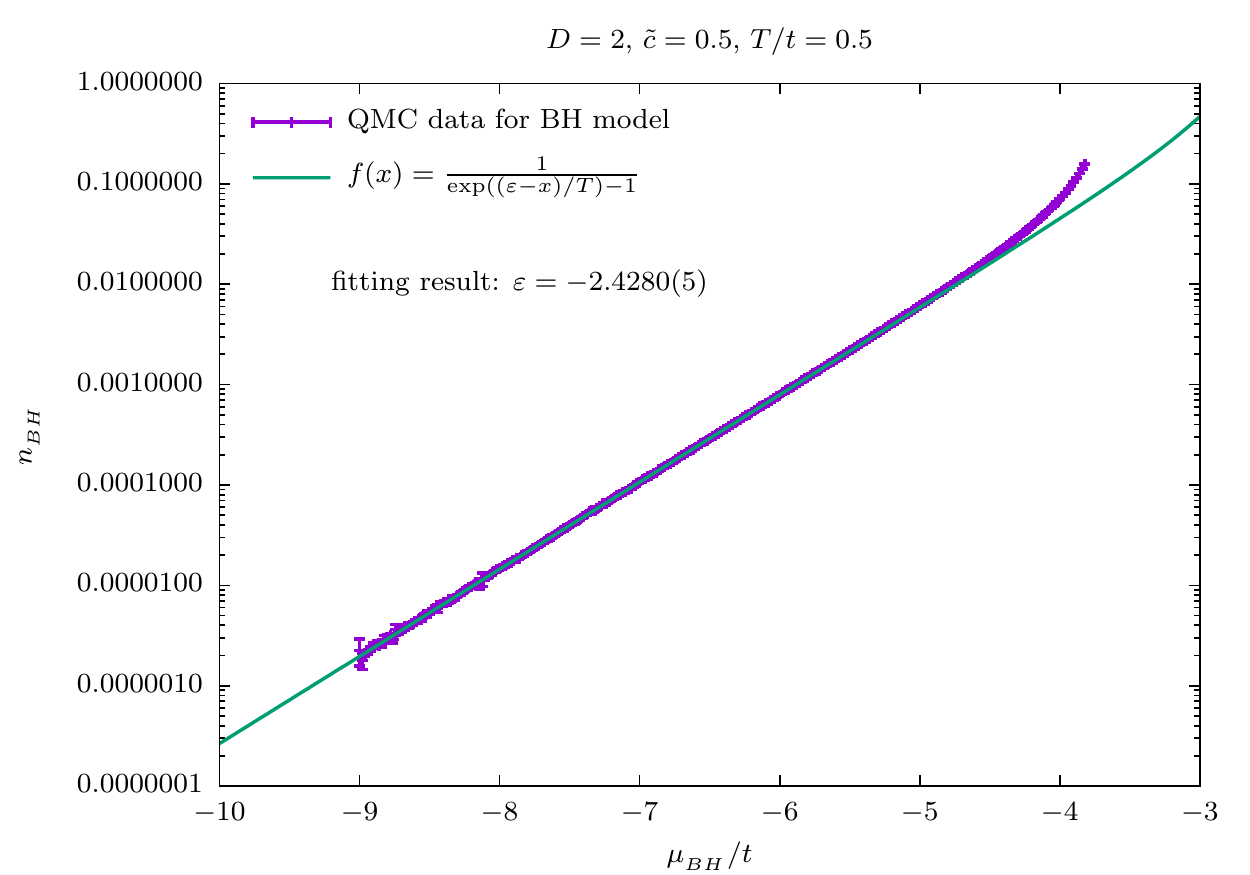}
	\caption{The particle number density in Bose-Hubbard model $n_{_{BH}}$ varies with the chemical potential $\mu_{_{BH}}/t$ at the specific parameters shown in the title of the figure. In order to calculate the pressure at, for example, $\mu_{_{BH}}/t=-4$, we measure $n_{_{BH}}$ in the region $\mu_{_{BH}}/t\in [-9,-4]$ with the interval $\Delta\mu_{_{BH}}/t=0.02$, and calculate the integral in Eq. \ref{eq:p_BH_mu_BH} numerically by trapezoidal rule and get $0.02686(6)$. We also fit the tail (here, we select $\mu_{_{BH}}/t\in [-9,-6.5]$) of the data by the formula Eq. \ref{eq:n_idealboson} with the fitting result $\varepsilon=-2.4280(5)$, and according to this, we obtain the estimate for the integral in the region $(-\infty,-9]$, which is around $1\times 10^{-6}$. Thus, in fact, the relative deviation induced by the truncation is of the order $10^{-5}$ which is pretty small compared with the relative statistical errors($\sim 10^{-3}$). We have taken account of the integral from $-\infty$ to $\mu_{_{BH}}^{min}$ in our results, but since $\mu_{_{BH}}^{min}$ we choose are small enough, we could safely ignore the effect brought by the truncation in principle.} \label{fig:n_mu}
\end{figure}

According to the Gibbs-Duhem equation,
\begin{eqnarray}
dp=nd\mu+sdT,
\end{eqnarray}
if the temperature is fixed, we have
\begin{eqnarray}
p(\mu)=p(\mu_0)+\int_{\mu_0}^{\mu}n(\mu')d\mu'. \label{eq:p_mu}
\end{eqnarray}
Considering that when $\mu\rightarrow -\infty$, the system will become vacuum with $n(\mu\rightarrow -\infty)=0$ and $p(\mu\rightarrow -\infty)=0$, we set $\mu_0=-\infty$ and Eq. \ref{eq:p_mu} becomes
\begin{eqnarray}
p(\mu)=\int_{-\infty}^{\mu}n(\mu')d\mu',
\end{eqnarray}
that is 
\begin{eqnarray}
p_{_{BH}}(\mu_{_{BH}})=\int_{-\infty}^{\mu_{_{BH}}}n_{_{BH}}(\mu')d\mu' \label{eq:p_BH_mu_BH}
\end{eqnarray}
for Bose-Hubbard model. During our calculation of this formula, we break the integral into two parts: one is integrating from a very small value, say $\mu_{_{BH}}^{min}$, to $\mu_{_{BH}}$, while the other is integrating in the region below $\mu_{_{BH}}^{min}$, that is $(-\infty,\mu_{_{BH}}^{min})$. Therefore, we first measure $n_{_{BH}}$ at different chemical potentials ranging from $\mu_{_{BH}}^{min}$ to $\mu_{_{BH}}$, with other parameters $U$, $t$ and $T$ keeping fixed, and then calculate the integral in Eq. \ref{eq:p_BH_mu_BH} numerically by trapezoidal rule in the region $[\mu_{_{BH}}^{min},\mu_{_{BH}}]$. And as for the region below $\mu_{_{BH}}^{min}$, we apply the distribution function for ideal Bosons
\begin{eqnarray}
n(\varepsilon,\mu,T)=\frac{1}{e^{(\varepsilon-\mu)/T}-1}, \label{eq:n_idealboson}
\end{eqnarray}
to fit the tail of our data for $n_{_{BH}}$ with the only fitting parameter $\varepsilon$, and based on the fitting result, we estimate the integral in the region $(-\infty,\mu_{_{BH}}^{min}]$. An example is shown in Fig. \ref{fig:n_mu}.

\begin{figure}[th]
	\centering
	\includegraphics[width=0.8\linewidth]{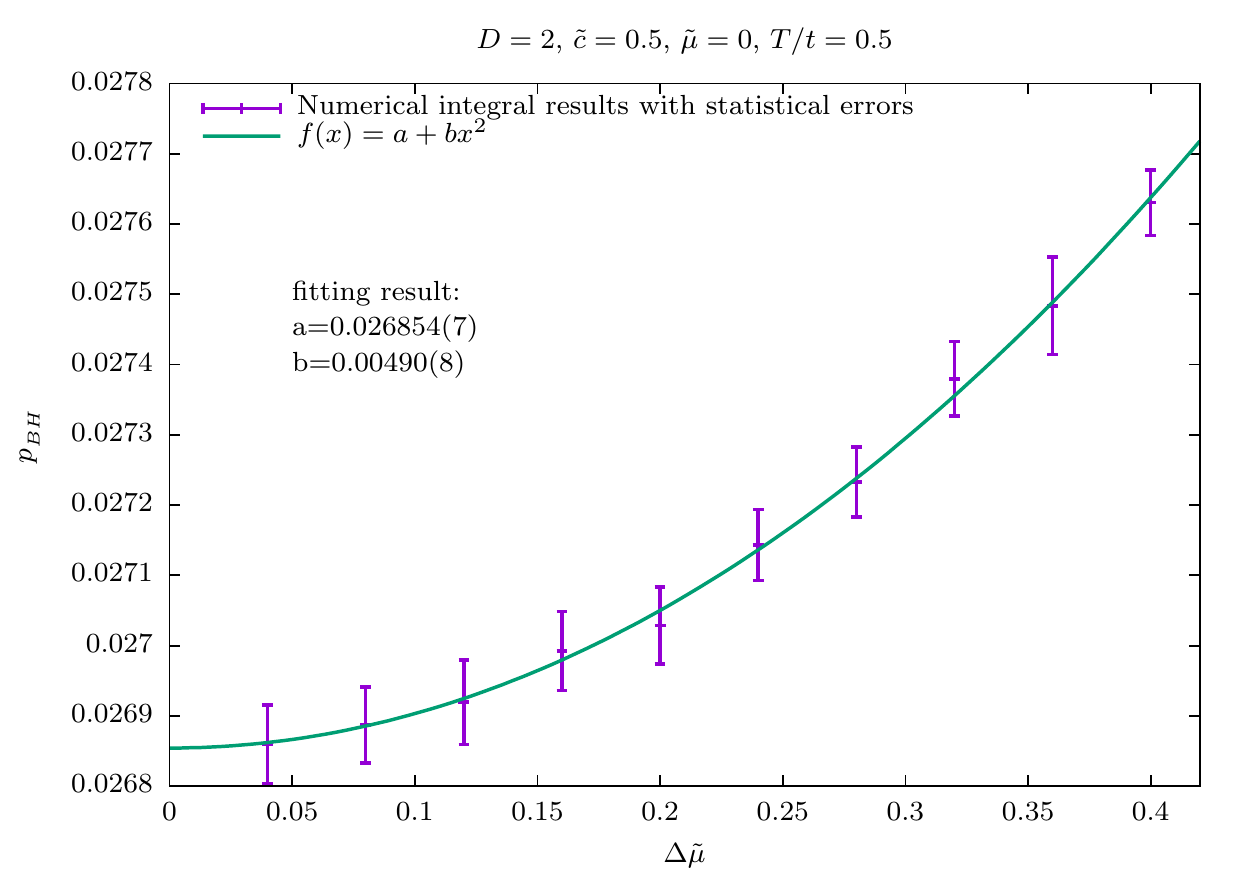}
	\caption{Numerical calculation results for the pressure in Bose-Hubbard model with different intervals $\Delta\tilde{\mu}$ during the numerical integrating. The error bars indicate the errors induced by the statistical uncertainty of the particle number density measured by QMC simulations. We can see that as the dimensionless interval $\Delta\tilde{\mu}$ decreases, the numerical integral results are approaching to a certain limit. This behavior could be well fitted by the formula $f(x)=a+bx^2$ as presented by the green line. According to the fitting result, the relative deviation resulting from the finite dimensionless interval is of the order $10^{-4}$ when $\Delta\tilde{\mu}=0.04$, whereas the relative statistical errors are typically of the order $10^{-3}$. Thus, under this circumstance, we could ignore the deviation brought by the finite interval during the numerical integrating.} \label{fig:p_dmu}
\end{figure}

To calculate the pressure at the critical point $p_{_{BH}}^c$, we typically set $\mu_{_{BH}}^{min}=\mu_{_{BH}}^c-10T$ and the interval for the numerical integral $\Delta\mu_{_{BH}}=0.04T$, where $\mu_{_{BH}}^c=-2Dt$ is the location of the critical point for the phase transition from a vacuum to a quantum liquid in Bose-Hubbard model. This corresponds to $\tilde{\mu}$ ranging from $-10$ to $0$ with the dimensionless interval $\Delta\tilde{\mu}=0.04$. As the example shown in Fig.~\ref{fig:p_dmu}, the errors coming from the finite interval during the numerical integrating can be ignored compared with the statistical errors.

In order to calculate the entropy of the system, we apply the following quasi-static process: keep all the parameters (including $t$, $U$ and $T$) fixed except the chemical potential $\mu$ varying from $\mu_0$ to $\mu$ extremely slowly, so that we could regard the system as always in equilibrium states. Following this process, we have
\begin{eqnarray}
S(\mu)=S(\mu_0)+\int_{\mu_0}^{\mu}\left.\frac{\partial S}{\partial\mu}\right|_{\mu=\mu'}d\mu'. \label{eq:S_mu}
\end{eqnarray}
On the other hand, the entropy of a system is defined by the following formula
\begin{eqnarray}
S=-Tr(\rho\ln\rho), \label{eq:S_rho}
\end{eqnarray}
where $\rho$ is the density matrix with the explicit expression
\begin{eqnarray}
\rho=e^{-\beta \mathcal{H}}/Z. \label{eq:rho}
\end{eqnarray}
Here, $Z=Tr\left(e^{-\beta \mathcal{H}}\right)$ is the partition function, $\mathcal{H}$ is the Hamiltonian in grand ensemble and typically could be written as $\mathcal{H}=\mathcal{H}_0-\mu N$. According to Eq. \ref{eq:rho}, we could rewrite Eq. \ref{eq:S_rho} in a more explicit form,
\begin{eqnarray}
S=-\beta\frac{\partial \ln Z}{\partial \beta}+\ln Z.
\end{eqnarray}
Therefore,
\begin{eqnarray}
\frac{\partial S}{\partial \mu}=-\beta\frac{\partial}{\partial \mu}\frac{\partial \ln Z}{\partial\beta}+\frac{\partial \ln Z}{\partial \mu}=\beta\left(\frac{\partial}{\partial\mu}\left\langle \mathcal{H} \right\rangle+\left\langle N \right\rangle\right).
\end{eqnarray}
Substitute this into Eq. \ref{eq:S_mu} and we can get
\begin{eqnarray}
S(\mu)&=&\int_{-\infty}^{\mu}\beta\left(\frac{\partial}{\partial \mu'}\left\langle \mathcal{H}(\mu') \right\rangle + \left\langle N(\mu') \right\rangle\right) d\mu' \nonumber\\
&=&\frac{1}{T}\left( \langle \mathcal{H}(\mu) \rangle + \int_{-\infty}^{\mu}\langle N(\mu') \rangle d\mu' \right) \nonumber\\
&=&\frac{V}{T}\left(\epsilon(\mu)+p(\mu)\right).
\end{eqnarray}
For Bose-Hubbard model, it becomes
\begin{eqnarray}
S_{_{BH}}(\mu_{_{BH}})=\frac{V}{T}\left[\epsilon_{_{BH}}(\mu_{_{BH}})+p_{_{BH}}(\mu_{_{BH}})\right],
\end{eqnarray}
where $\epsilon_{_{BH}}(\mu_{_{BH}})$ could be directly measured in our QMC simulations and $p_{_{BH}}(\mu_{_{BH}})$ could be obtained via the formula Eq.~\ref{eq:p_BH_mu_BH} by numerical integral techniques.

\section{Comparing simulations with existing experimental measurements}

For our theory and simulation results in both 1D and 2D systems, we compare them with existing experiments, as shown in Figs.~2 $\sim$ 4 in the main text. Below we briefly review the sources of experimental data and our re-analysis for some of them.

For the 1D gas experiment  by the Kaiserslautern group, we directly obtain the $S_{\mathrm{c}}/N$ and $\tilde{n}_{\mathrm{c}}$ data from Ref.~\cite{Ott13PRA}. For the 1D gas experiment by the USTC group~\cite{GuanYuan17PRL}, we obtain the original data of equation of state ($n(\mu)$) that are taken under multiple temperatures, and  re-analyze them to obtain the critical pressure $p_{\mathrm{c}}$ and the critical entropy density $s_{\mathrm{c}}$ based on $s = \left(\frac{\partial P}{\partial T}\right)_{\mu}$.

For the 2D gas experiment by the ENS, we obtain the $S_{\mathrm{c}}/N$ data directly from Ref.~\cite{Dalibard11PRL}. For the 2D gas experiments by the Chicago group~\cite{Hung11nature, Zhang12Science} where the 2D  lattice gases / continuous-space gases  are experimentally observed to satisfy scale invariance, we either obtain data from these two references or process the original data to extract the needed thermodynamic observables.

For another experiment by the Chicago group~\cite{Ha13PRL},  we obtain the $\tilde{n}_{\mathrm{c}}$ and $\tilde{p}_{\mathrm{c}}$ data from Ref.~\cite{Ha13PRL}, but do not have sufficient data to extract $S_{\mathrm{c}}/N$ because the authors neither test scale invariance in the strong coupling regime nor perform groups of measurements for the same $\tilde{c}_{\mathrm{2D}}$ under multiple temperatures.

Finally, while our theoretical and numerical data agree quite well with existing experiments~\cite{Ott13PRA, GuanYuan17PRL, Zhang12Science, Dalibard11PRL, Hung11nature, Ha13PRL} in Figs.~2, 3, and 4 of the main text, we note that there are experimental factors that in principle  can lead to the break-down of scale invariance to some extent in experimental systems -- in particular, in the strongly interacting experimental systems. For example, three-body loss effects scale as the fourth power of the atomic scattering length. In addition, when one prepare strongly interacting experimental systems by approaching  the unitarity limit (Feshbach resonance) or using deep optical lattices, the scattering amplitude at finite temperatures can become momentum-dependent. These and other experimental factors could in principle cause the break-down of scale invariance to some extent in experimental systems, but the quantitative characterization of such break-down still remains sparse. The comparison of our theoretical and numerical data with existing experiments provide new insights regarding this issue. The overall good agreements suggest that in existing experiments, a description based on scale invariance is fairly consistent with the underlying physics within the current level of experimental uncertainties. The residual small discrepancies between our results and existing experiments may come from practical factors including inelastic losses and finite-temperature effects in experiments.

\bibliography{intFESref}

\end{document}